\documentclass[12pt, preprint]{aastex}
\slugcomment{Submitted to Astrophysical Journal}

\newcommand{\nuc}[2]{${}^{#2} \rm #1$}

\def\gtaprx {\lower .14ex\hbox{\rlap{\raise .9ex\hbox{\hskip .3ex
	{\ifmmode{\scriptscriptstyle >}\else
		{$\scriptscriptstyle >$}\fi}}}
	\kern -.4ex{\ifmmode{\scriptscriptstyle \sim}\else
		{$\scriptscriptstyle\sim$}\fi}}}
\def\ltaprx {\lower .14ex\hbox{\rlap{\raise .9ex\hbox{\hskip .3ex
	{\ifmmode{\scriptscriptstyle <}\else
		{$\scriptscriptstyle <$}\fi}}}
	\kern -.4ex{\ifmmode{\scriptscriptstyle \sim}\else
		{$\scriptscriptstyle\sim$}\fi}}}

\newcommand{\g}{ {\rm g} }
\newcommand{\cm}{ {\rm cm} }
\newcommand{\s}{ \,{\rm s} }
\newcommand{\K}{ {\rm K} }

\newcommand{\km}{{ \rm km}}
\newcommand{\ms}{{ \rm ms}}

\newcommand{\psec}{{ \rm s}^{-1}}
\newcommand{\gpccm}{{ \rm g \,\, cm^{-3}}}
\newcommand{\ergps}{\rm ergs \,\, {s}^{-1}}

\newcommand{\del}[2]%
{\frac{\mathrm{d}{#2}}{\mathrm{d}{#1}}}
\newcommand{\Del}[2]%
{\frac{\mathrm{D}{#2}}{\mathrm{D}{#1}}}
\newcommand{\ddel}[2]%
{\frac{\mathrm{d}^2{#2}}{\mathrm{d}{#1}^2}}

\newcommand{\pdel}[2]%
{\frac{\partial{#2}}{\partial{#1}}}
\newcommand{\pddel}[2]%
{\frac{\partial^2{#2}}{\partial{#1}^2}}

\newcommand{\laplace}{\bigtriangleup}

\renewcommand{\vec}[1]{\mathbf{#1}}

\newcommand{\Ms}{M_{\odot}}
\newcommand{\DMs}{M_{\odot}\rm \,s^{-1}}

\shorttitle{MHD simulations of collape of a massive star}
\shortauthors{Fujimoto et al.}

\begin{document}

\title{
Magnetohydrodynamic Simulations of A Rotating Massive Star Collapsing to A Black Hole}

\author{
Shin-ichirou Fujimoto\altaffilmark{1},
Kei Kotake\altaffilmark{2},
Shoichi Yamada\altaffilmark{2,3},
Masa-aki Hashimoto\altaffilmark{4},
and Katsuhiko Sato\altaffilmark{5,6}}

\altaffiltext{1}{Department of Electronic Control, 
Kumamoto National College of Technology, Kumamoto 861-1102, Japan;
fujimoto@ec.knct.ac.jp}

\altaffiltext{2}{\textit{Science \& Engineering, Waseda University,
3-4-1 Okubo, Shinjuku, Tokyo 169-8555, Japan}}

\altaffiltext{3}{\textit{Advanced Research Institute for Science \&
Engineering, Waseda University, 3-4-1 Okubo,
Shinjuku, Tokyo 169-8555, Japan}}

\altaffiltext{4}{
Department of Physics, School of Sciences, 
Kyushu University, Fukuoka 810-8560, Japan}

\altaffiltext{5}{
Department of Physics,
School of Science, the University of Tokyo, 7-3-1 Hongo,
Bunkyo-ku,Tokyo 113-0033, Japan}

\altaffiltext{6}{
Research Center for the Early Universe,
School of Science, the University of Tokyo,7-3-1 Hongo,
Bunkyo-ku, Tokyo 113-0033, Japan}


\begin{abstract}
We perform two-dimensional, axisymmetric, 
magnetohydrodynamic simulations of 
the collapse of a rotating star of 40 $\Ms$ and 
in the light of the collapsar model of gamma-ray burst.
Considering two distributions of angular momentum, 
up to $\sim 10^{17} \rm cm^2\, \psec$, 
and the uniform vertical magnetic field, 
we investigate the formation of an accretion disk around a black hole
and the jet production near the hole. 
After material reaches to the black hole with the high angular momentum, 
the disk is formed inside a surface of weak shock.
The disk becomes in a quasi-steady state for stars 
whose magnetic field is less than $10^{10}$ G before the collapse.
We find that the jet can be driven by the magnetic fields
even if the central core does not rotate 
as rapidly as previously assumed and outer layers of the star 
has sufficiently high angular momentum.
The magnetic fields are chiefly amplified inside the disk
due to the compression and the wrapping of the field.
The fields inside the disk propagate to the polar region 
along the inner boundary near the black hole through the Alfv{\'e}n wave,
and eventually drive the jet.
The quasi-steady disk is not an advection-dominated disk
but a neutrino cooling-dominated one.
Mass accretion rates in the disks are greater than
$0.01 \DMs$ with large fluctuations.
The disk is transparent for neutrinos.
The dense part of the disk, which locates near the hole, 
emits neutrino efficiently at a constant rate
of $<$ 8$\times 10^{51} \ergps$.
The neutrino luminosity is much smaller than those from supernovae 
after the neutrino burst.
\end{abstract}




\section{Introduction}

\setcounter{footnote}{0}


During the collapse of a massive star greater than 35-40 $\Ms$,  
stellar core is considered to promptly collapse to a black hole~\citep{ww95,heger03}.
Stellar material greater than several solar masses 
may fall onto the hole at extremely high accretion rates 
($> 1 \DMs$)~\citep{ww95,heger03}.
If the star has sufficiently high angular momentum before the
collapse, an accretion disk is likely to be formed around the hole~\citep{mw99}.
Jets are suggested to be launched from the inner region of the accretion disk
near the hole through magnetic and/or neutrino processes.
Gamma-ray bursts (GRBs) are expected to be driven by the jets.
This scenario of GRBs is called a collapsar model~\citep{w93}.
Assisted by the accumulating observations implying the association between 
GRBs and the death of massive stars (e.g. \cite{galama98,hjorth03,zkh04}), 
the collapsar model seems most promising.


In the light of the collapsar scenario, 
\cite{mw99} and \cite{mwh01} have performed 
two-dimensional hydrodynamic simulations of 
rotating massive stars during collapse after the formation of a black hole.
Taking account of the viscous heating with
$\alpha$-prescription~\citep{ss73} and the neutrino cooling, 
they showed that an accretion disk is formed around the hole 
if the progenitor has a sufficient range of angular momentum.
The disk cools via advection chiefly and 
the structure of the disk is shown to be well described with 
a steady, one-dimensional model of 
an advection dominated accretion flow (ADAF)~\citep{pwf99}.
However, jets cannot be produced in their hydrodynamic simulations.
They suggested that collimated outflows can produce through 
magnetohydrodynamic (MHD) and/or neutrino processes~\citep{mw99, mwh01}.
Jet propagation through stellar envelope has been investigated via 
hydrodynamic simulations~\citep{aloy00,aloy03,zwm03}.
The jets are shown to be ultrarelativistic and collimated in the envelope,
though the jets are assumed to be initiated from the central region of 
the collapsing star and the the thermal energy is deposited at a rate about
$10^{51} \ergps$ at the region.

In order to investigate jet production in a collapsar,
\cite{proga03} have performed MHD simulation of stellar collapse
of a rapidly rotating 25$\Ms$ progenitor, 
whose iron core is assumed to a 1.7$\Ms$ black hole and
magnetic field to be radial (monopole-like) and uniform.
They considered neutrino cooling processes in an optically thin regime 
and resistive heating, whose properties are highly uncertain.
In their relatively short ($\le 0.28 \s$) simulation 
of a part of the collapsar (about 10km - 5000km), 
relativistic jets up to $0.2c$ 
are revealed to be launched and collimated magnetically.
\citet{mizuno04a} also have shown that
relativistic jets ($\le 0.3c$) are formed
in general relativistic MHD calculations of a collapsing massive star of 15 $\Ms$.
Faster jets can be produced near a black hole
with larger rotational parameters~\citep{mizuno04b}.
It should be emphasized that their simulations are performed 
with very small computational domain of 360km$\times$360km and 
for very short duration about $5 \ms$ for a 3$\Ms$ black hole.
We note that
MHD calculation have been performed for a collapsing star with 
relatively small mass in the previous studies~\citep{proga03,mizuno04a,mizuno04b}, 
though the formation of a black hole is assumed.

On the other hand, 
jet production via neutrino annihilation is examined with 
a one-dimensional disc model~\citep{pwf99}.
\citet{mw99} have estimated the energy deposition rate due to neutrino 
annihilation to be $\sim 5\times 10^{50}$ ergs$\rm \, s^{-1}$
using the disc model~\citep{pwf99} and the mass accretion rates 
obtained from their hydrodynamic simulations.
However, the deposition rate highly depends on the structure of the disks.
Disks related to GRBs cannot be ADAFs as in \citet{mw99,pwf99,proga03}, 
but they are a neutrino-cooling dominated flow (NDAF) 
if neutrino-cooling is more efficient than advective cooling~\citep{npk01}. 
Properties of NDAF are extensively investigated using one-dimensional, 
height integrated disk models with detailed micro physics~\citep{km02,yoko04}
and two-dimensional simulations~\citep{lee05}.
The disks could be convection-dominated flows~\citep{npk01}.
Which type of an accretion disk is realized in GRBs could depend
on the size of the disk and mass accretion rates through the outer
boundary of the disk~\citep{npk01}.

In the present paper, 
we perform Newtonian MHD simulations of a collapsing massive star of 40$\Ms$
to investigate the formation of an accretion disk and the production of 
jets from the disk in a collapsar.
We consider magnetized stars with both rapidly and slowly rotating
cores, which are assumed to collapse to a black hole promptly.
Our simulations covers from iron core to an oxygen-rich layer 
of the collapsing star, or 50km-10 000km, and 
are performed for a long term up to $\sim 4 \s$, which is much longer
than the previous studies~\citep{proga03,mizuno04a,mizuno04b}, 
to examine long term evolution of the collapsing star, in particular, 
properties of the disk and jets during the late phase of the collapse.
GRBs related to the final stage of a massive star have long duration 
$> 2 \s$.
Therefore such long calculation is important for understanding
the relation between long GRBs and the disk/jet system formed in the
collapsing star.

In \S 2 we present basic equations of MHD calculation of the collapsing
star, a numerical code, and initial conditions of the star.
Model parameters and numerical results are shown in \S 3.
We discuss our numerical grid to check resolving 
magneto-rotational instability (MRI)
and compare our results with those obtained 
in previous works~\citep{proga03,mizuno04a} in \S 4. 
Finally, we summarize our results in \S 5.

\section{Numerical method and Initial conditions}

We carry out Newtonian MHD calculation of 
the collapse of a rotating massive star of 40$\Ms$.
In the present study, the core of the star is assumed to be 
collapsed to a black hole promptly, 
although the prompt formation of the black hole depends on
not only the progenitor mass (or the core mass) but also
an equation of state (EOS)
and the angular velocity distribution inside the core
~\citep[e.g.][]{seki04,seki05}.
Calculation is performed over the region from 50km to 10 000km of the star.
Fluid is freely absorbed through the inner boundary of 50km, 
which mimics a surface of the black hole.
The black hole mass, $M$, is initially set to be that of 
the central region of the progenitor $\le 50\km$ (0.001$\Ms$) 
and is continuously increased by the mass of 
the infalling gas through the inner boundary at $r_{\rm in}$, 
\begin{equation}
 \Delta M = \Delta t 4\pi r_{\rm in}^2 \int^{\pi/2}_0 \rho v_r \sin\theta \, d\theta,
\label{eq:M}
\end{equation}
during time step $\Delta t$, 
where $\rho$ and $v_r$ are the density and radial velocity 
estimated at the inner boundary, respectively.

General relativistic hydrodynamic simulations show that
black holes are formed at 0.1-$0.2 \s$ after the onset of collapse~\citep{seki05}.
In our models, the core is absorbed through the inner boundary
and the black hole mass reaches about 2 $\Ms$
at the freefall time, $\sqrt{3\pi/32{\rm G}\rho_c} \sim 0.1 \s$, 
after the onset of the collapse, as we shall see later.
Here $\rho_c$ is the core density of $\sim 10^{10} \gpccm$.
Hence, our procedure can approximate
the collapsing phase to a black hole in an appropriate level.
We note that 
\citet{proga03} have performed an MHD simulation after the formation of 
a black hole of 1.7$\Ms$.

\subsection{Input Physics and Numerical Code}

To calculate the structure and evolution of the collapsing star, 
we solve the Newtonian MHD equations, 
\begin{equation}
\Del{t}{\rho}+\rho\nabla\cdot\vec{v}=0,
\end{equation}
\begin{equation}
\rho \Del{t}{\vec{v}}=-\nabla P -\rho \nabla (\Phi +\Psi)
+\frac{1}{4\pi}\left(\vec{\nabla}\times\vec{B}\right)\times\vec{B},
\end{equation}
\begin{equation}
\rho\Del{t}{}\left(\frac{e}{\rho}\right)=-P\nabla \cdot \vec{v}
-L_{\nu},
\end{equation}
\begin{equation}
\pdel{t}{\vec{B}}= \vec{\nabla} \times \left(\vec{v}\times 
\vec{B}\right)\label{eq:1},
\end{equation}
\begin{equation}
\laplace{\Phi} = 4\pi G \rho,
\end{equation}
where $\rho,\vec{v},P,\Phi,\Psi,\vec{B},e$, and $L_{\nu}$ 
are the mass density, the fluid velocity, 
the pressure other than the magnetic pressure,
the gravitational potential of fluid,
the gravitational potential of the central object,
the magnetic field, the internal energy density, 
and the neutrino cooling rate, respectively.
We denote the Lagrange derivative as $D/Dt$.

The numerical code for the MHD calculation
employed in this paper is based on the ZEUS-2D code~\citep{sn92}.
We have extended the code to include
a realistic EOS~\citep{kotake04a} based on the 
relativistic mean field theory~\citep{shen98}.
For lower density regime ($\rho < 10^5 \gpccm$), 
where no data is available in the EOS table with the Shen EOS, 
we use another EOS, which includes contributions
from an ideal gas of nuclei, radiation, and electrons and positrons
with arbitrary degrees of degeneracy~\citep{bdn96}.
We carefully connect two EOS at $\rho = 10^5 \gpccm$
for physical quantities to vary continuous in density at a given temperature.

We consider neutrino cooling through 
electron-positron pair capture on nuclei,
electron-positron pair annihilation, and
nucleon-nucleon bremsstrahlung.
The total neutrino cooling rate is evaluated with 
a simplified neutrino transfer model based on 
the two-stream approximation~\citep{dpn02}, 
with which we can treat the optically thin and thick regimes
on neutrino reaction approximately.
We ignore resistive heating, whose properties are highly uncertain,
not as in \citet{proga03}.
We note that the change in the electron fraction, $Y_e$, is ignored
in the MHD calculation (or, $DY_e/Dt = 0$).

Spherical coordinates, $(r, \theta, \phi)$ are used 
in our simulations and the computational domain is extended over
$50 \km \le r \le 10 000 \km$ and $0 \le \theta \le \pi/2$
and covered with 200($r$) $\times$ 24($\theta$) meshes, 
with which we can resolve a fastest growing mode of MRI for
most models (see details in \S 4.1).
The location of the inner boundary is greater than that in 
\citet{proga03} ($\sim 10$ km).
We discuss how the location of the inner boundary affects 
our numerical results later (\S \ref{sec:inner-boundary}).
We fix ratios of the size of the meshes as
$\Delta r_{k+1}/\Delta r_{k} = 1.02$ ($\Delta r \ge 1 \times 10^6$ cm) 
and $\Delta \theta_{k+1}/\Delta \theta_{k} = 1$.
We assume the fluid is axisymmetric and the mirror symmetry 
on the equatorial plane.
We mimic strong gravity around the black hole in terms of 
the pseudo-Newtonian potential~\citep{pw80}: 
\begin{equation}
  \Psi =  - \frac{GM}{r -r_{\rm g}},
\end{equation}
where $G$ is the gravitational constant and
$r_{\rm g} = 2GM/c^2$ is the Schwarzschild radius.

\subsection{Initial conditions}

We set the initial profiles of the density, temperature, internal energy
density, and electron fraction 
to those of the spherical model of a 40$\Ms$ massive star 
before the collapse~\citep{hashimoto95}.
The radial and azimuthal velocities are set to be zero initially, 
and increase due to the collapse induced by the central hole and 
self-gravity of the star.
The computational domain is extended from the iron core to an inner
oxygen layer.
The star of about 4$\Ms$ is enclosed with the computational domain.
The boundaries of the silicon layers between 
the iron core and the oxygen layers are located 
at about 1800 km (1.88$\Ms$) and 3900km (2.4$\Ms$), respectively.
We adopt an analytical form of the angular velocity 
$\Omega(r)$ of the star before the collapse:
\begin{equation}
 \Omega(r) = \Omega_0 \frac{R_0^2}{r^2 + R_0^2},
 \label{eq:omega0}
\end{equation}
as in the previous study 
of collapsars~\citep{mizuno04a, mizuno04b} and SNe~\citep{kotake04a, yama04a}.
Here $\Omega_0$ and $R_0$ are parameters of our model.
We consider two sets of ($\Omega_0, \, R_0$); 
($10\,\rm rad \, s^{-1}$, 1000km) (case with rapidly rotating core) 
and 
($0.5\,\rm rad \, s^{-1}$, 5000km) (case with slowly rotating core).
For these sets of $\Omega_0$ and $R_0$, 
the maximum specific angular momentum is about 
$10^{17}\,\cm^2 \rm \, s^{-1}$, 
which is comparable to that of the Keplerian motion 
around a black hole of 3$\Ms$ at 50km.
Therefore the centrifugal force can be larger than 
the gravitational force of the central black hole
and the formation of a disk-like structure is expected near the hole.
Figure 1 shows the specific angular momentum distribution on the equatorial plane
for two cases and for the previous works \citep{mw99, proga03}.
The angular momentum for the two cases are comparable 
near the outer boundary of the computational domain.
We find that the specific angular momentum adopted in \citet{mw99} 
is similar to those for the model with
$\Omega_0 = 10\,\rm rad \, s^{-1}$ and $R_0 = 1000$ km.


Initial magnetic field is assumed to be uniform, 
parallel to the rotational axis, and weak elsewhere
($\beta = 8\pi P/B^2 \gg 1$).
We consider cases with the initial magnetic field, 
$B_0 = 10^8, 10^{10}$ and $10^{12}$ G.
It should be noted that 
the magnetic pressure is much smaller than the other pressure
even if $B_0$ is equal to $10^{12}$ G.
The models are labeled by R8, R10, R12, S8, S10, and S12, 
in which the first character, R(rapidly rotating core) or S(slowly rotating core), 
indicates the set of $\Omega_0$ and $R_0$, 
and the numeral, 8, 10, and 12, equals $\log_{10} (B_0/\rm G)$.

\section{Numerical Results}

We summarize model parameters and features in Table 1.
Columns (2) through (4) give model parameters, 
with which initial conditions are represented for each model.
Column (5) gives the time at the end of each run, $t = t_f$, 
after the onset of collapse, at which $t$ is equal to 0.
Column (10) gives the the mass of the black hole at $t = t_f$.
Simulations for models with $B_0 = 10^{8}$, $10^{10}$, and, $10^{12}$ G
are performed for $3 \times 10^5$, $2 \times 10^5$, and $1 \times 10^5$ 
time steps, respectively.
Columns (6) and (7) ((8) and (9)) give
ratios of the rotational (magnetic) to the potential energy integrated
over the computational domain at $t = 0$ and $t = t_f$, respectively.

As we shall see later, jets are produced near the black hole 
for R10, R12, S10, and S12.
Properties of the jets are summarized in Table 2.
Column (2) gives $t_{\rm jet}$ when jets are passed through 1000km.
Columns (3), (4), (5), and (6) represent 
the mass, magnetic, kinetic, and internal energies of the jets,
respectively. 
The mass and energies of the jets are defined as those ejected
through the outer boundary.

\subsection{Collapse of a star with a rapidly rotating core}\label{sec:fast-core}

\subsubsection{MHD features}\label{sec:fast-core-mhd}


We present magnetohydrodynamic features of the collapsing star
for R10 as a representative case.
Figure \ref{fig:r10-contours} shows contours of density (left panels)
and those of the ratio of $P_{\rm mag}$ to the pressure (right panels) 
for R10 at $t = $ 1.6625, 2.0193, and, $2.6225 \s$.
Initially, the star collapses nearly spherical 
and becomes aspherical gradually as the gas accretes 
with higher angular momentum. 
When the gas with $\sim 10^{17} \cm^2\rm \,s^{-1}$ 
falls the black hole ($\sim 0.1 \s$),
weak shock appears near the remnant 
and propagates outward slowly (top and middle left panels).
The propagation velocity of the shock is several thousands km $\rm s^{-1}$, 
which is much larger than the Alfv{\'e}n velocity and 
slightly less than the sound velocity.
A disk-like structure forms inside the shock (top left panel). 
We find that after the formation of the disk, 
the flow consists of the three parts;
(i) the nearly spherical inflow outside the shock, 
(ii) the quasi-steady, equilibrium disk,
and (iii) the inflow along the disk surface.
The disk is in quasi-steady state and supported by the centrifugal force mainly
and the gas pressure, which is larger than the magnetic pressure. 
The magnetic pressure is enhanced inside the disk 
(top and middle right panels) and finally
drives a jet, whose expansion velocity reaches 0.1c
near the black hole along the rotational axis 
as we can see later (Figure \ref{fig:r10-jet}).
In this region the density drops below $10^7 \gpccm$ (bottom left panel), 
the radial and toroidal components of the magnetic field are $10^{14}$ G, 
and the magnetic pressure is greater than the gas pressure (bottom left panel).

We pay attention to a region near the inner boundary (300km $\times$ 300km) 
to investigate the generation of jets.
We can clearly see the phase of jet production in Figure \ref{fig:r10-jet},
in which contours of $P_{\rm mag}$ for R10 at $t = $ 2.5301, 2.5340, 
and, $2.5368\s$ (from left to right) are presented.
The magnetic fields are amplified inside the disk and
propagates just above the inner boundary to the polar region
via the Alfv{\'e}n wave (left and center panels).
The wave is enough fast ($\sim$10 000 km $\rm s^{-1}$) to propagate along the boundary
before the infall to the black hole along with the accreting gas.
The jets are eventually driven by the magnetic pressure near the boundary
(center panel) and ejected along the rotational axis (right panel).
In this way, the magnetic energy, which is enhanced inside the disk, 
can be transported to the jets.

We calculate another model with the same angular velocity distribution
as R10 but without magnetic field, or R0, to examine the cause of the weak
shock.
We find that a weak shock also appears in this non-magnetized model. 
Therefore, the shock seems not to be originated from any magnetic
processes.
It is possibly generated via centrifugal barrier.
We also note that similar accretion shock have appeared 
in a hydrodynamic simulation of a collapsar~\citep{mw99}.

It is pointed out that the magnetic field is exponentially amplified 
through MRI in an accretion disk~\citep{bh98}.
Similar amplification of the magnetic field is also suggested to occur
in a rotating star during collapse 
from one-dimensional calculations~\citep{wmw02, awml03}.
We examine the field amplification in our two-dimensional models of
a rotating massive star.
Figure \ref{fig:r10-Emag} shows the time evolution of 
the magnetic fields integrated over the computational domain. 
We denote $\int (B_i^2/8\pi) dV$ as $E_{m,\,i}$ ($i = r, \theta, \phi$).
The initial magnetic field is vertical so that
the toroidal component of the magnetic field vanishes 
and $E_{m,\,r} = E_{m,\,\theta}$ initially.
After the accretion of high angular momentum material ($t \ge 0.1 \s$),
the radial and azimuthal components of the magnetic field 
are amplified exponentially in time and the toroidal component is 
rapidly enhanced.
We note that the MRI of the vertical magnetic field produces 
the $X$ component of the magnetic field and
the wrapping of the poloidal field generates the toroidal field.
We find that the amplification of the radial and azimuthal fields 
is mainly due to compression of the field initially and MRI in the later phase
and that the toroidal field can be amplified by wrapping of 
the seed poloidal field.
In fact, 
the radial and azimuthal fields are amplified as $\exp(t/0.1\s)$, 
as shown in Figure \ref{fig:r10-Emag},
and the growth timescale of MRI, $\tau_{\rm MRI}$, is $\sim 0.1\s$.
On the other hands, the characteristic timescale 
for the field wrapping, $\tau_{\rm WRAP}$, is calculated as 
\begin{eqnarray}
\tau_{\rm WRAP}
&=& 4\pi\left|\frac{\pdel{t}{B_{\phi}}}{B_{\phi}}\right|^{-1} \nonumber \\
&=& 4\pi \left|
\frac{B_{X}}{B_{\phi}}\left(X\pdel{X}{\Omega}\right)+
\frac{B_{Z}}{B_{\phi}}\left(X\pdel{Z}{\Omega}\right)
\right|^{-1} \label{eq:tau_wrap}
\end{eqnarray}
~\citep{taki04}.
Here $B_{X}$ and $B_{Z}$ are the $X$ and $Z$-components of the magnetic
field, respectively.
The toroidal magnetic field is therefore rapidly enhanced 
through the field wrapping if the poloidal field dominates over
the toroidal one and the gas is differentially rotating.
This is the case in R10 at $t \sim 0.1 \s$.
We note that $\tau_{\rm WRAP}$ is shorter than $\tau_{\rm MRI}$
during the rapid amplification phase of the toroidal field
($t \sim 0.1\s$) as in a magnetized core collapse SN~\citep{taki04}.
As the toroidal field increases, $\tau_{\rm WRAP}$ becomes longer 
(eq. (\ref{eq:tau_wrap})). 
After the toroidal field becomes comparable to the poloidal field
($t > 0.1 \s$), the growth timescale of the toroidal magnetic field 
is comparable to that of the poloidal field.
Note that $\tau_{\rm MRI}$ ($\sim 0.1 \s$) becomes comparable or 
shorter than $\tau_{\rm wrap}$ in this phase.
Because our two-dimensional axisymmetric simulations cannot follow 
the amplification of the toroidal magnetic field due to MRI, 
the toroidal field may be saturated as a lower level.
At the end of the simulation, 
the total magnetic energy, kinetic, and rotational energies
are 0.57 $\times 10^{51}$, 1.52 $\times 10^{51}$, and 
1.50 $\times 10^{51}$ ergs, respectively.

For higher initial magnetic field of the progenitor, 
the magnetic field is more rapidly amplified than R10.
Figure \ref{fig:r12-contours} shows contours of density (left panels)
and those of the ratio of $P_{\rm mag}$ to the pressure (right panels) 
for R12 at $t = $ 0.2378 and $0.3560 \s$.
A jet launches by the magnetic pressure (right top panel)
just after the formation of a disk (left top panel)
near the central remnant along the rotational axis.
In the polar region, 
the radial and toroidal components of the magnetic field
are comparable and reach $10^{15}$ G just before the launch of the jet.
It should be emphasized that
the density of the jet in R12 is much higher than that of R10.
Figure \ref{fig:r12-jet-tscale} shows various time scale in jets.
Solid, dashed, and, dotted lines denote 
the ejection time, the neutrino cooling time through pair capture, 
and, that via pair annihilation, respectively.
We find that the ejection time is much smaller than the neutrino cooling
time and that jets cannot be cooled through neutrino.

For lower initial magnetic field, hydrodynamic features are similar to R10.
For R8, a disk-like structure is formed inside a shock surface as in R10.
The disk is in quasi-steady state and
is supported by the centrifugal force and the pressure gradient.
The radial velocity is lower than 
the rotational velocity, which has a Keplerian profile, 
by about three orders of magnitude.
The magnetic pressure is lower than the gas pressure elsewhere, 
and the toroidal component of the field attains to $10^{14}$ G.

\subsubsection{Disk properties}\label{sec:fast-core-disk}


To examine properties of the disk, 
we present physical quantities 
near the equatorial plane ($\theta = 88.1^\circ$)
in Figure \ref{fig:r10-disk}, for R10 at $1.6625 \s$, 
at which the weak shock propagates near 2000km
(top left panel in Figure \ref{fig:r10-contours}). 
The radial velocity is lower than the near-Keplerian rotational velocity
by about two orders of magnitude.
The disk is in equilibrium and mainly supported by the centrifugal
force, which is greater than the pressure gradient near the black hole.
Inside the shock surface, 
the flow is highly convective (bottom left panel), 
which is clearly seen in Figure \ref{fig:r10-inner-region}.
Here contours of the ratio of $P_{\rm mag}$ to the pressure are shown 
at an inner region (500km$\times$500km) for R10 at $t = 1.6625 \s$
(top right panel in Figure \ref{fig:r10-contours}).
We find that inside the disk, 
velocity field is highly complicated so that
the magnetic field has also rather complex structure.

Figure \ref{fig:r10-q_nu} shows contour of logarithmic
specific neutrino cooling rate 
through pair capture on nuclei, for R10 at $1.6625 \s$, defined as 
\begin{equation}
q_{\rm cap} = 9.2 \times 10^{23} \rho X_{\rm nuc} 
\left(\frac{T}{10^{11} \rm K}\right)^6 \, \ergps\,cm^{-3},
\label{eq:qcap} 
\end{equation}
where $X_{\rm nuc}$ is the mass fraction of nucleons and 
is approximated with 
\begin{equation}
 X_{\rm nuc} = 8.2 \times 10^8 
\frac{T_{\rm MeV}^{9/8}}{\rho^{3/4}} \exp(-7.074/T_{\rm MeV}),
\label{eq:xnuc}
\end{equation}
or unity, whichever is smaller with 
the temperature in MeV, $T_{\rm MeV}$~\citep{wb92}. 
Density contours of $10^8, 10^9, 10^{10}$, and $10^{11}$ $\gpccm$ are
plotted on Figure \ref{fig:r10-q_nu}.
We find that 
neutrino cooling is efficient inside the shock, where the temperature 
is higher than $10^{10} \K$ via shock heating, 
and much efficient near the equatorial plane of the disk 
with high densities 
because of more efficient pair capture in denser regions.
Neutrino luminosity stays at a constant level about 
$5 \times 10^{51}$ ergs $\rm s^{-1}$, as we shall see later (Figure \ref{fig:Lnu}).
We find that neutrino cooling via e$^\pm$ capture dominates over
those through electron-positron pair annihilation and
nucleon-nucleon bremsstrahlung
by one or two orders of magnitude inside the disk
(bottom panels in Figure \ref{fig:r10-disk-pl}).

At an inner region of disks, 
neutrino cooling time via e$^\pm$ capture, $e/q_{\rm cap}$, 
which is 0.1-1 $\s$, is comparable or 
smaller than accretion time, $|r/v_r|$
(bottom panels in Figure \ref{fig:r10-disk-pl}).
Thus, the disks cool not advection but neutrino emission in the region.
Indeed, we find that the density and temperature profiles are well described 
with those of {\itshape not ADAF but NDAF}, or, 
$\rho \sim r^{-51/20}$ and $T \sim r^{-3/10}$ 
(top panels in Figure \ref{fig:r10-disk-pl}).
For ADAF, the profiles have different power laws, or, 
$\rho \sim r^{-3/2}$ and $T \sim r^{-5/8}$~\citep[e.g.,][]{fujimoto00}. 
We note that the gas pressure is dominated over 
the radiation and degenerate pressure in our models
other than outer part of the disks, where
the radiation pressure is comparable to the gas pressure
(bottom right panel in Figure \ref{fig:r10-disk}).
The magnetic pressure is lower than the pressure
other than the region near the rotational axis where 
the toroidal component of the field is amplified 
to $10^{15}$ G (top right panel in Figure \ref{fig:r10-disk}).

In order to examine whether the disk is opaque to neutrino or not,
we calculate the height $h_{\nu}$ at which the neutrino optical depth
$\tau_\nu(R)$ is equal to 2/3, 
\begin{equation}
 \tau_\nu(R) = \int^{h_{\rm max}(R)}_{h_{\nu}(R)} \rho(R,z) \kappa_\nu(R, z) dz = \frac{2}{3},
\end{equation}
as in \citet{sm03}.
Here $h_{\rm max}$ and $\kappa_\nu$ are 
the height of the computational domain from the equatorial plane 
and the neutrino opacity at the distance from the rotational axis, $R$.
We note that the optical depth is simply set to be the sum of the depths for
e, $\mu$, and $\tau$ neutrinos.
The neutrino decoupling surface is found to be torus-like and 
very limited to the inner region ($\le$ 100km) 
of the disk near the black hole if exists.
For $t \ge 0.3 \s$, the surface is located at the region
below 30km from the equatorial plane near the hole.
We find that almost disk is therefore optically thin to neutrino, 
and that neutrino is hardly trapped inside the disk.


Figure \ref{fig:r10-Mdot} shows the time evolution of 
the accretion rate through the inner boundary at 50km, defined as, 
\begin{equation}
 \dot{M} = 4\pi r^2 \int^{\pi/2}_0 \rho v_r \sin\theta \, d\theta,\label{fig:S10}
\end{equation}
where $\rho$ and $v_r$ are estimated at the boundary.
The rates through parts of the boundary, 
$\dot{M}_{\rm pol}$ ($\theta \le 20^\circ$) and
$\dot{M}_{\rm disk}$ ($\theta \ge 50^\circ$), 
are also presented in Figure \ref{fig:r10-Mdot}.
At the beginning of the collapse, 
material with low angular momentum falls through the inner boundary
at very high accretion rates ($\ge 10 \DMs$)
for the freefall time of core ($\sim 0.1 \s$). 
As gas accretes onto the black hole with high angular momentum ($t \ge 0.1 \s$),
the polar infall dominates over the disk accretion.
$\dot{M}_{\rm disk}$ is much lower than $\dot{M}$, which is 
comparable with $\dot{M}_{\rm pol}$, by one order of magnitude.
During quasi steady collapse ( $t > 1.0 \s$), 
$\dot{M}$ is $\sim$ 0.2-0.5 $\DMs$ with fast fluctuations and 
gradual decrease. 
$\dot{M}_{\rm disk}$ varies between $0.02 \DMs$ and $0.07 \DMs$
in this phase and sharply decreases after the ejection of the jet.

\subsection{Collapse of a star with a slowly rotating core}\label{sec:slow-core}

\begin{figure} 
 \plotone{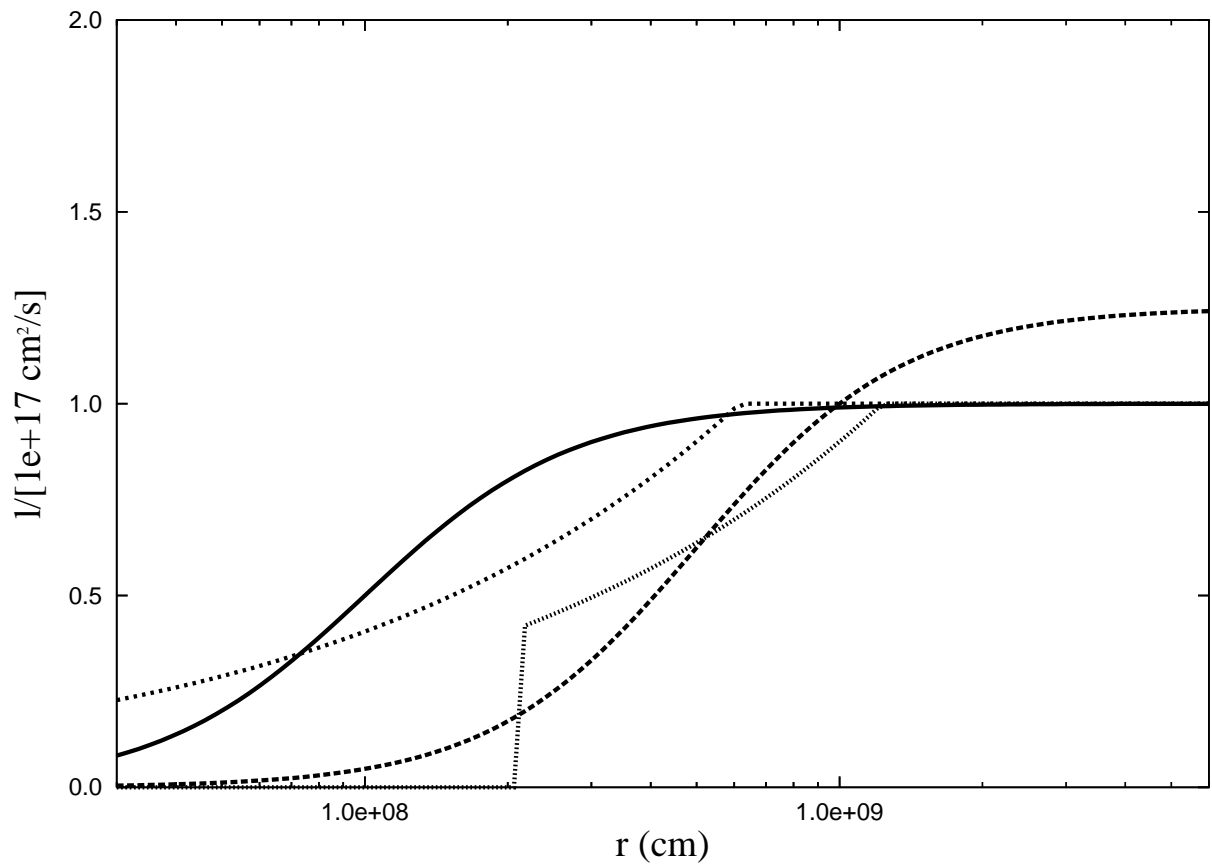}
 \caption{
Initial angular momentum distribution inside the collapsing star
at the equatorial plane.
The solid, dashed, dotted, and dash-dotted lines represent distributions 
for rapidly rotating core ($R_0 = $ 1000km and 
$\Omega_0 = $ 10 rad$\rm \, s^{-1}$) and 
for slowly rotating core ($R_0 = $ 5000km and 
$\Omega_0 = $ 0.5 rad $\rm \, s^{-1}$), 
those of \citet{mw99} and \citet{proga03}, respectively.} 
\label{fig:init-ang-momentum}
\end{figure}
\begin{figure}
\epsscale{.8}
\plottwo{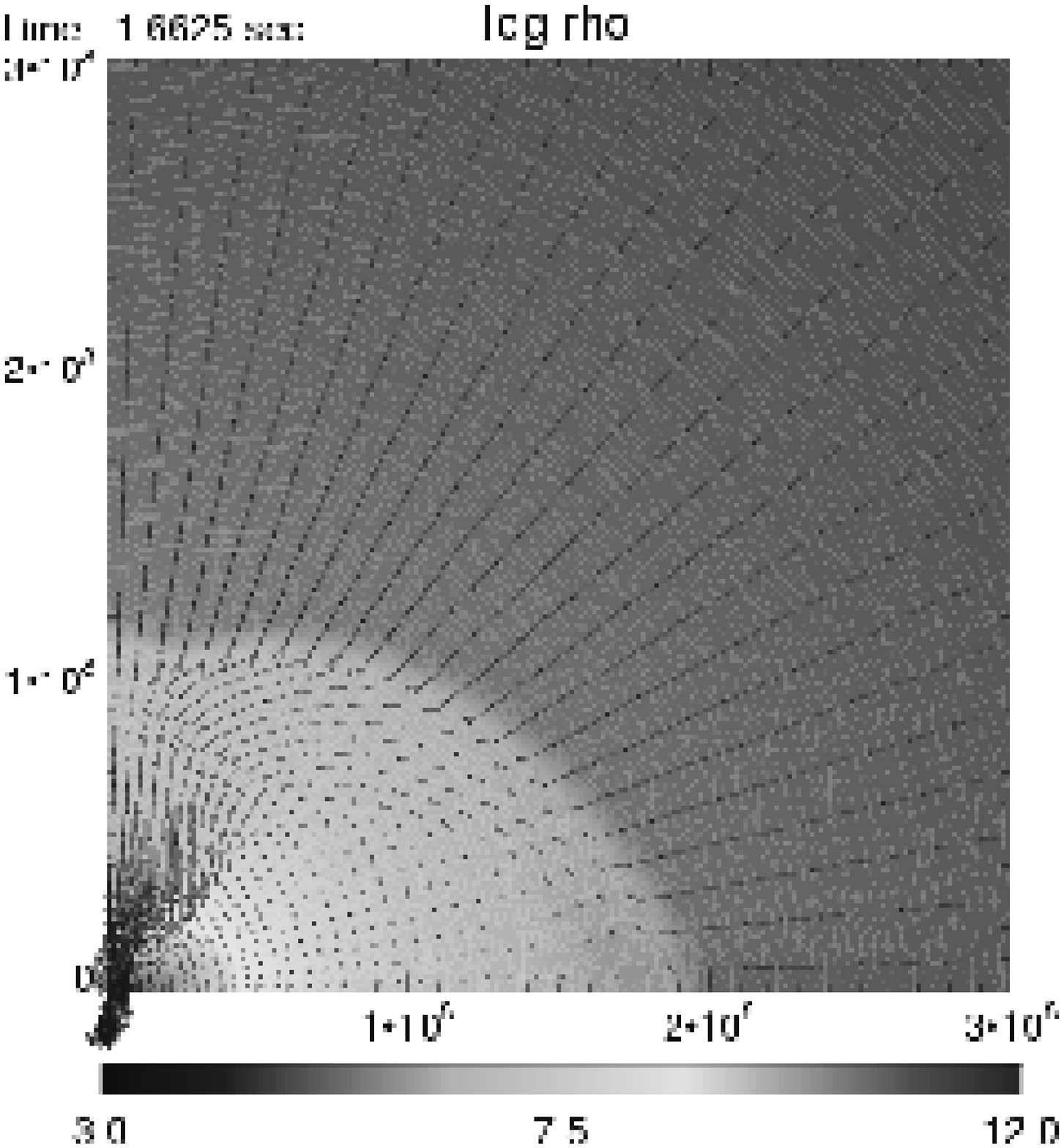}{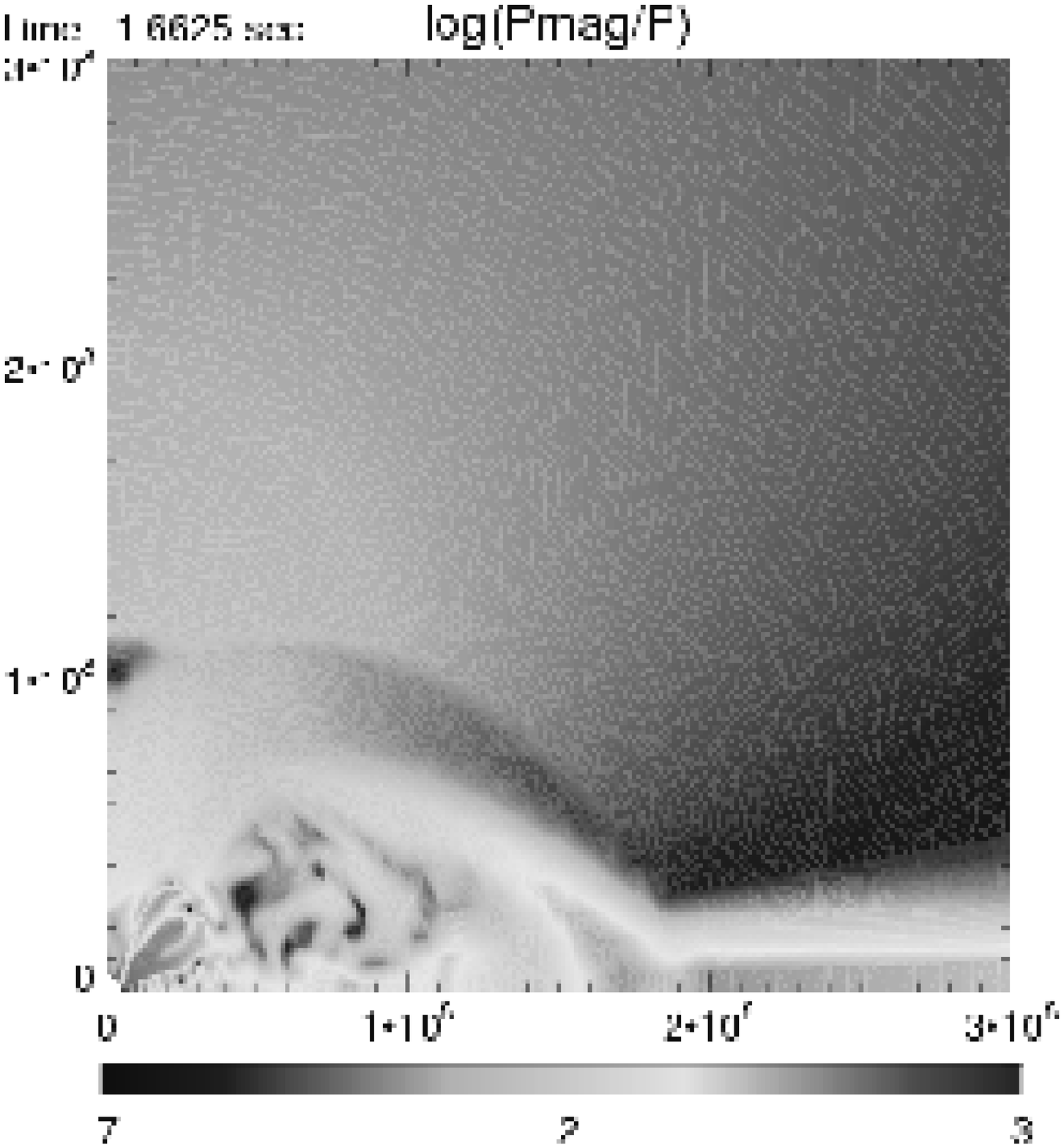}
\plottwo{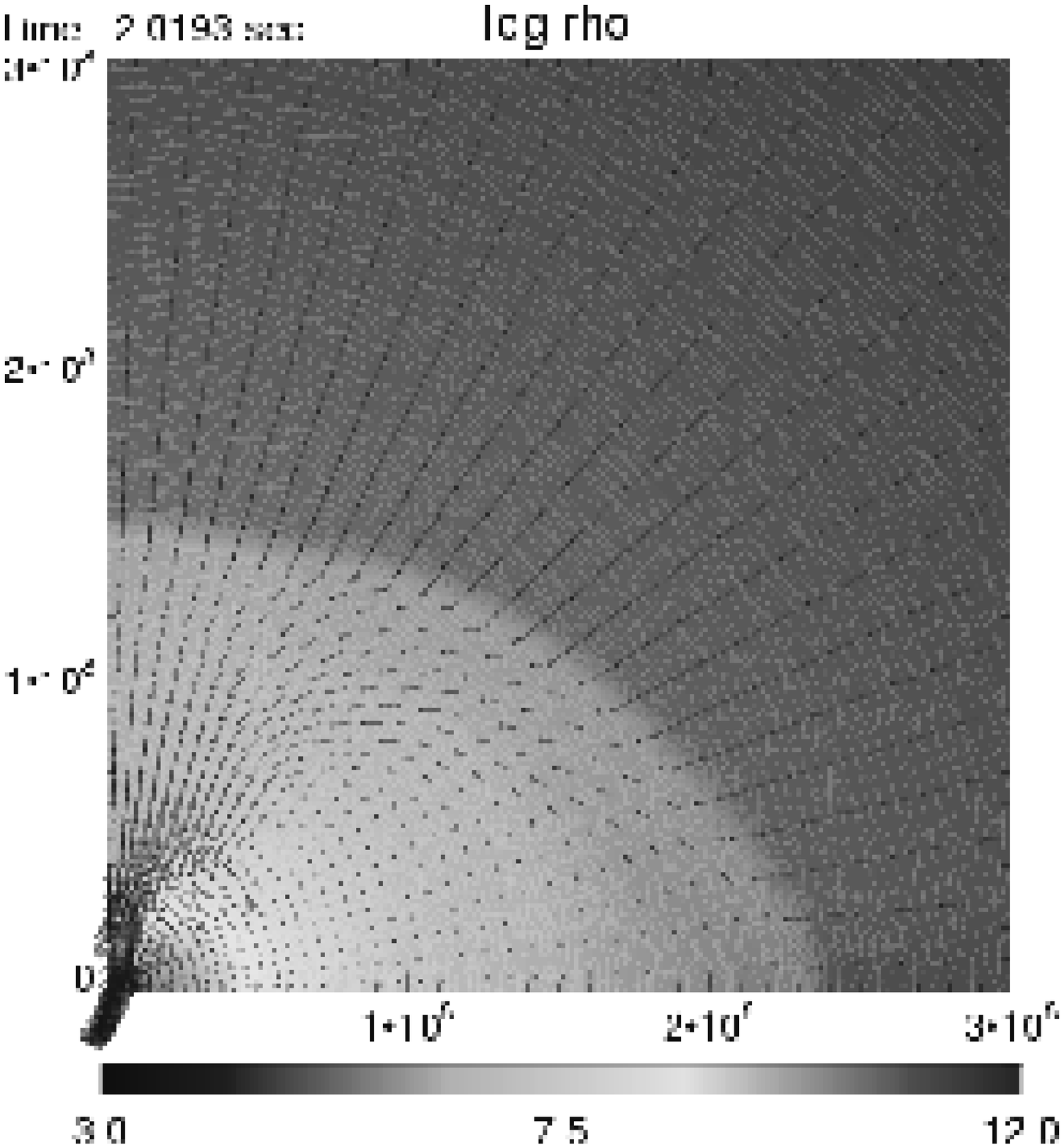}{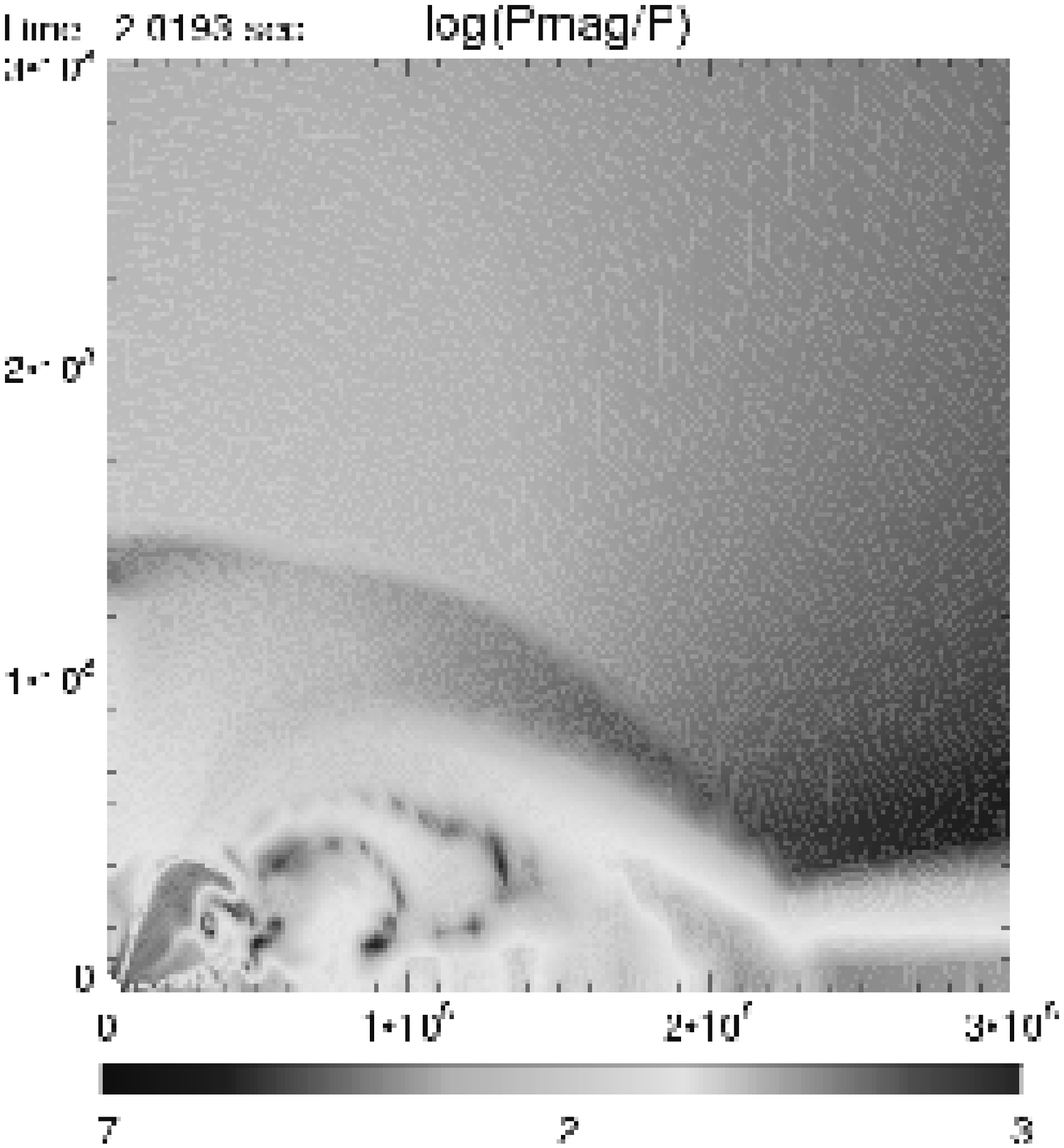}
\plottwo{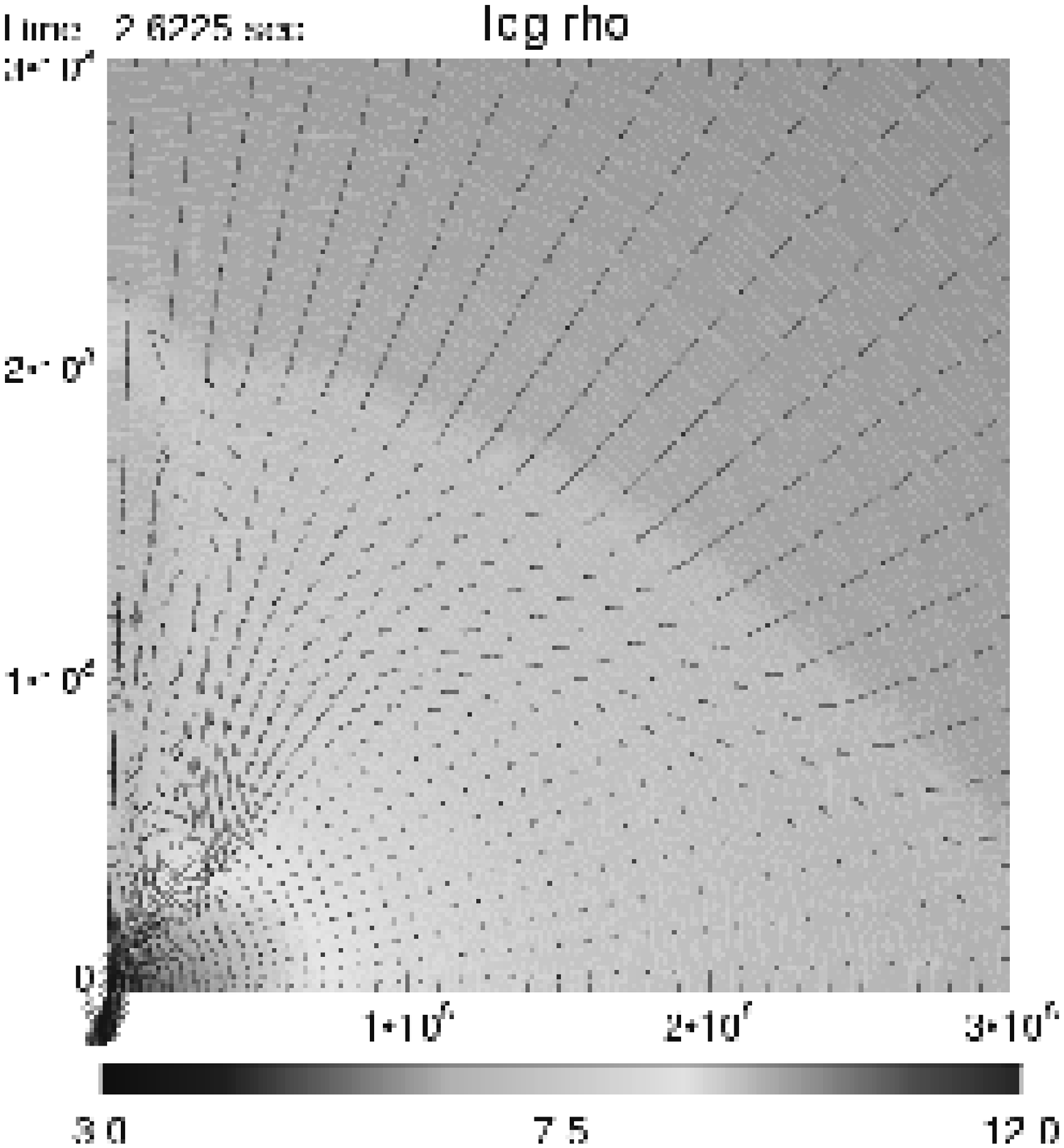}{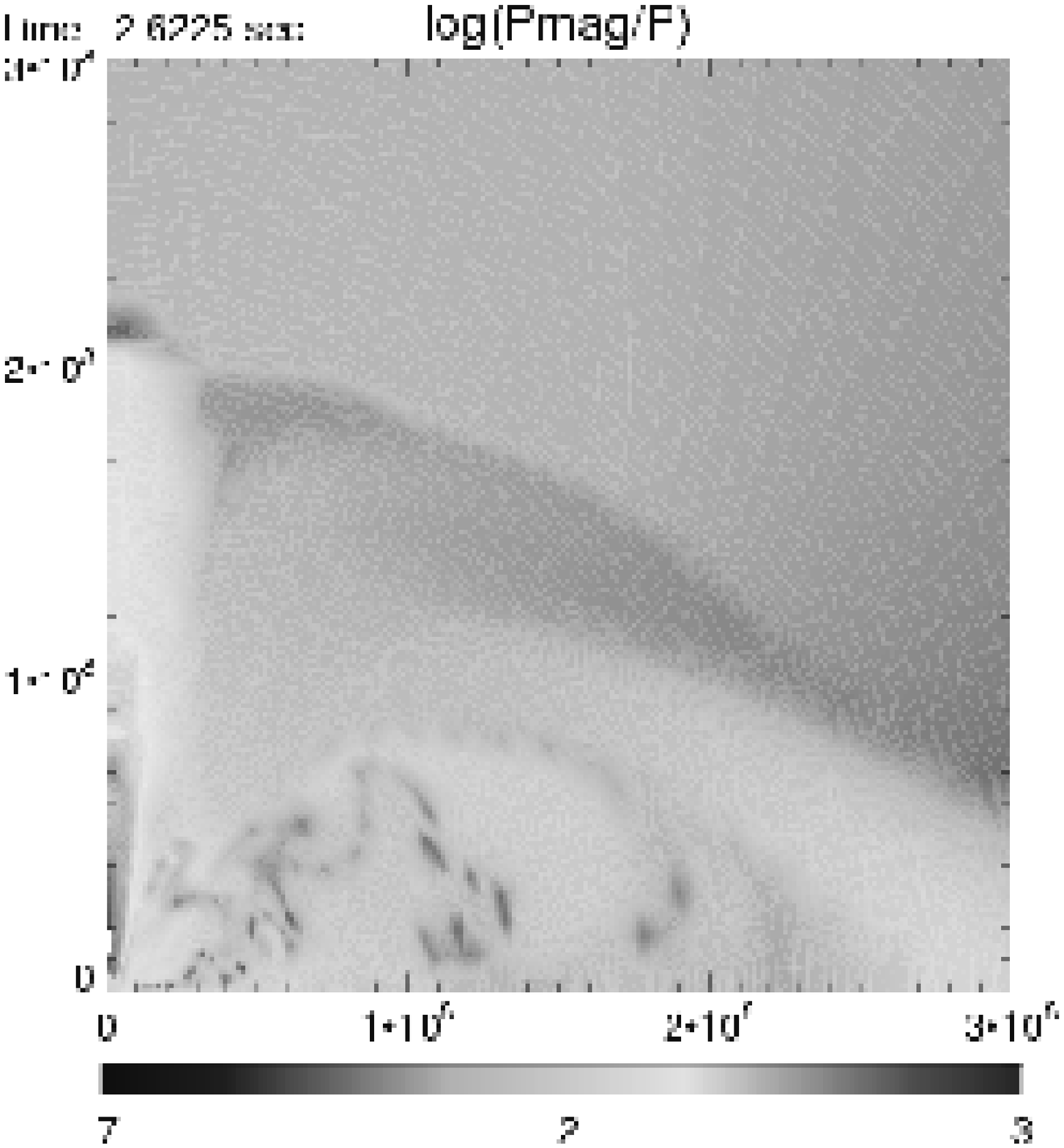}
\epsscale{1}
\caption{
Contours of the density (left panels) and 
the ratio of $P_{\rm mag}$ to the pressure (right panels) 
for R10 at $t = 1.6625$, 2.0193, and, $2.6225\s$ (from top to bottom).
} \label{fig:r10-contours}
\end{figure}
\begin{figure}
\epsscale{.3}
\plotone{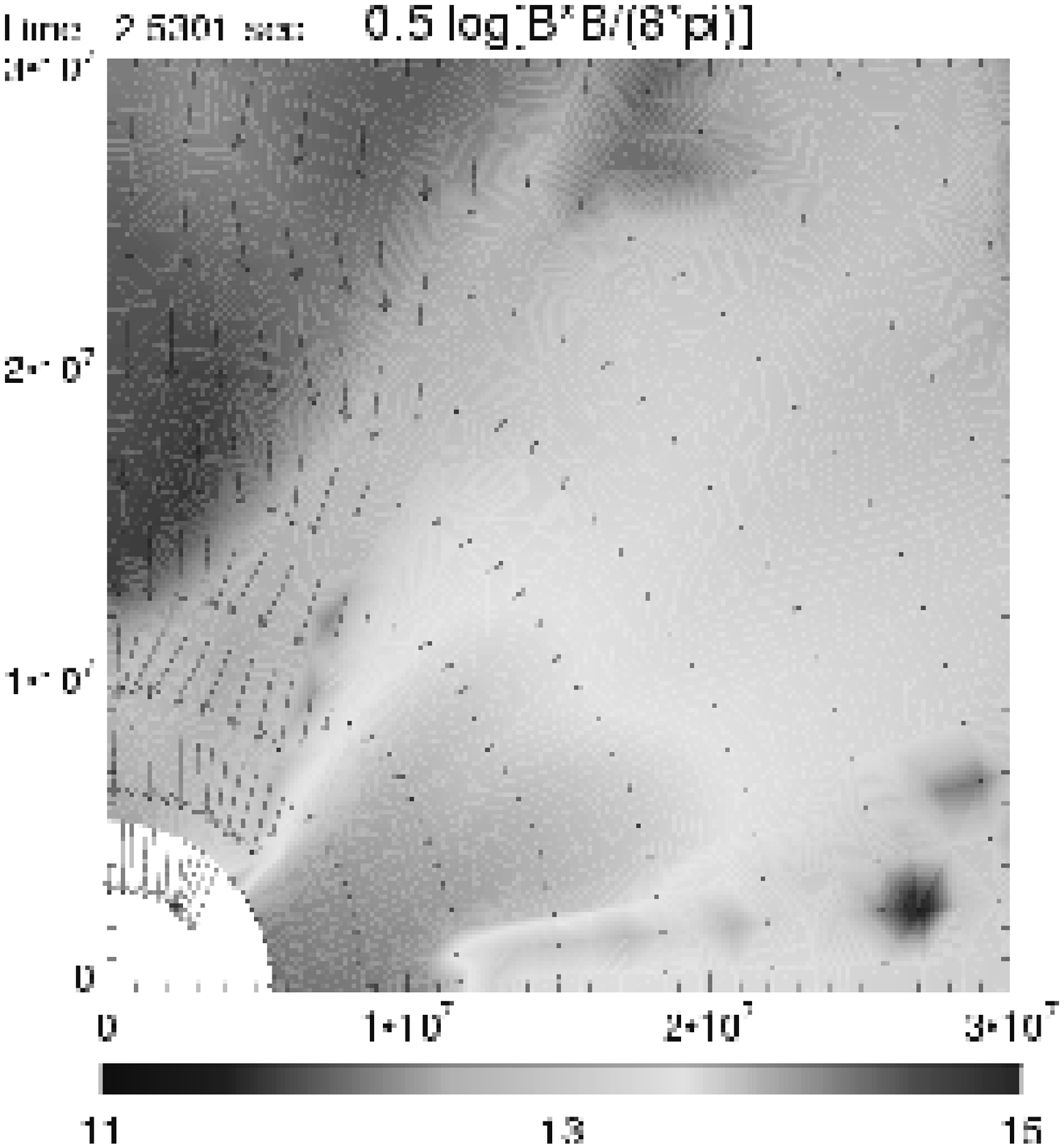}   
\plotone{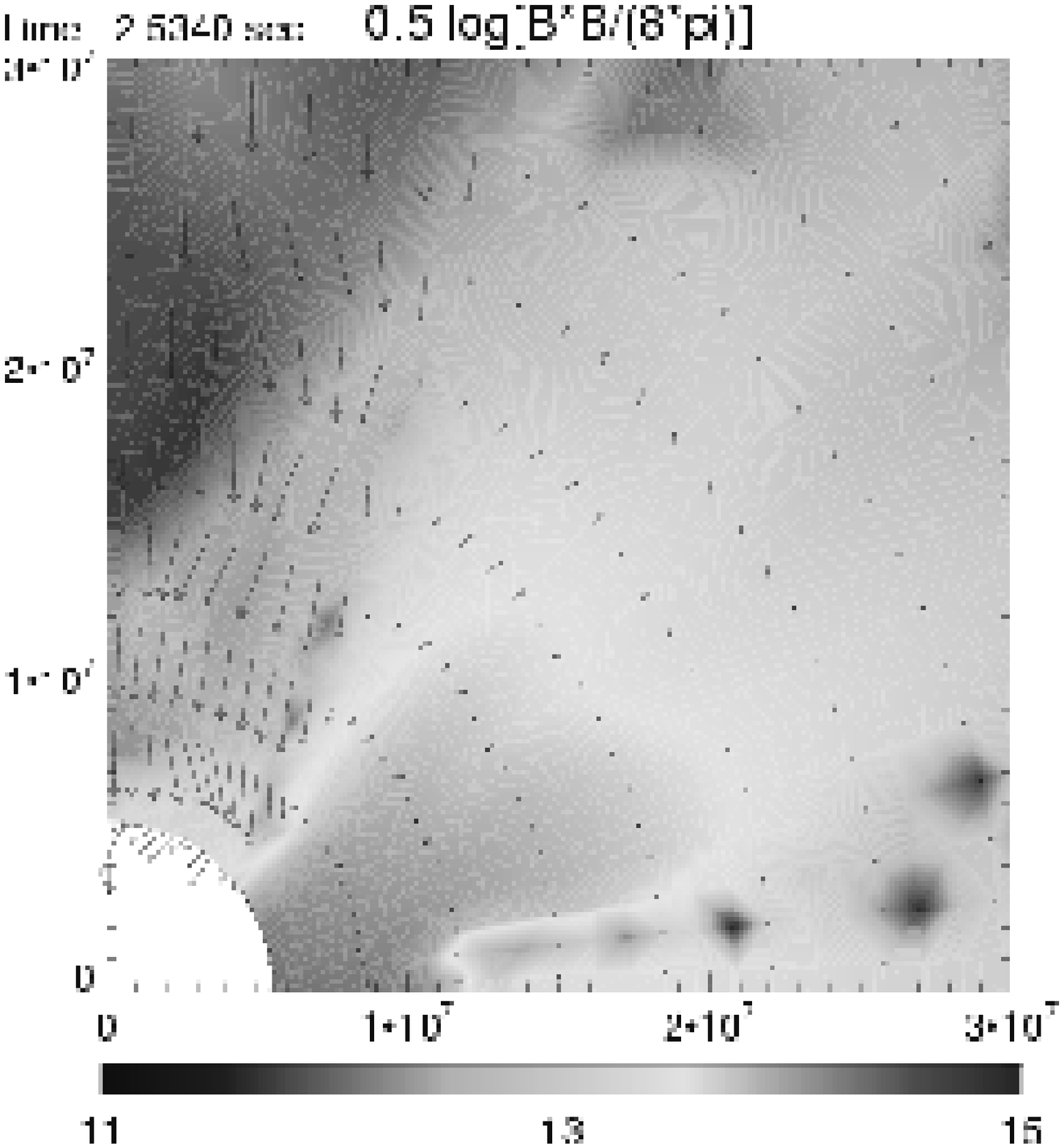}   
\plotone{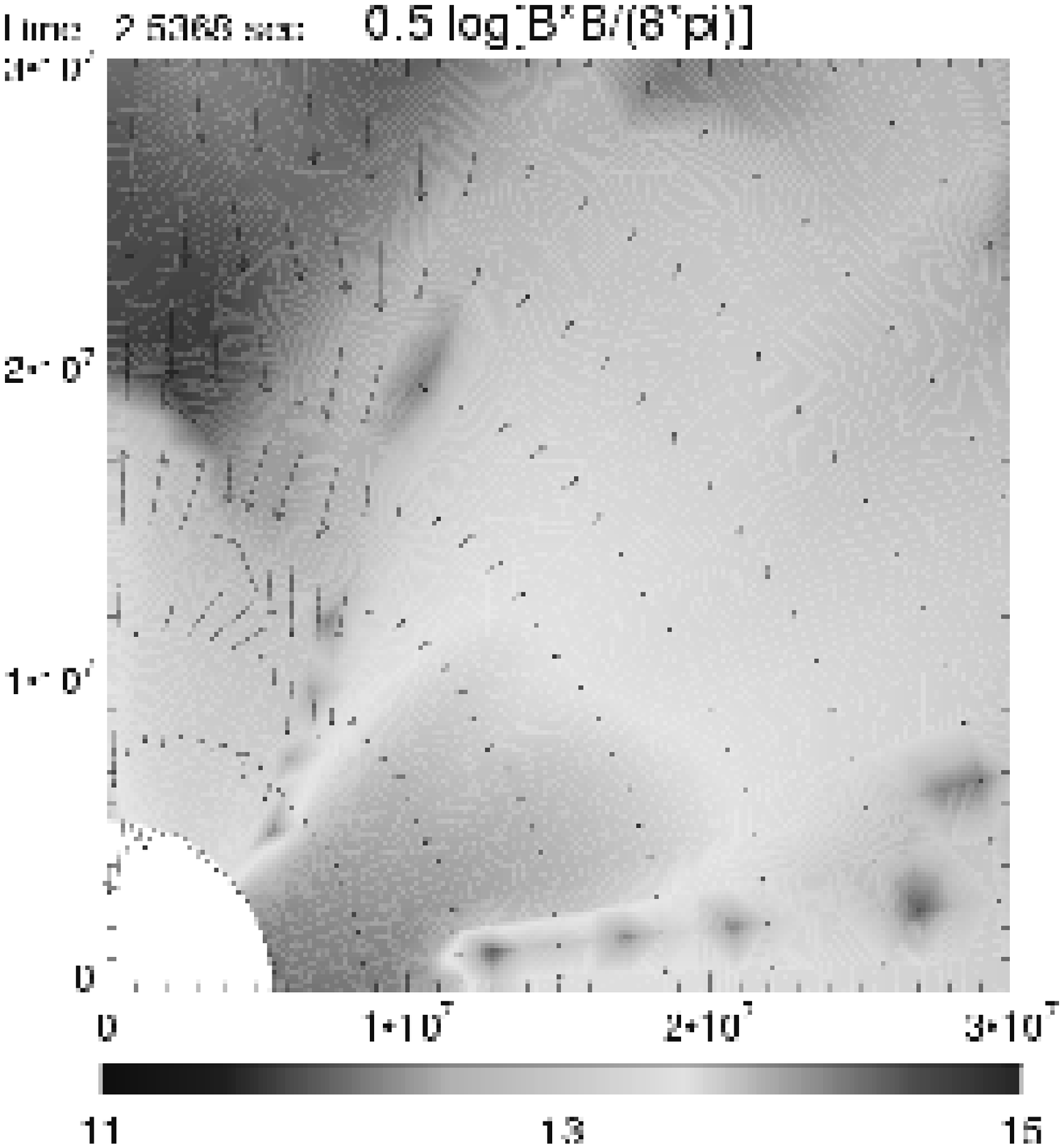}   
\epsscale{1}
\caption{
Contours of $P_{\rm mag}$ 
for R10 at $t = 2.5301$, 2.5340, and $2.5368\s$ (from left to right).
Magnetic field enhanced inside the disk propagates along the inner 
boundary (left and center panels).
Jets are driven by the magnetic pressure near the boundary
(center panel) and propagates along the rotational axis (right panel).
} \label{fig:r10-jet}
\end{figure}
\begin{figure}
 \plotone{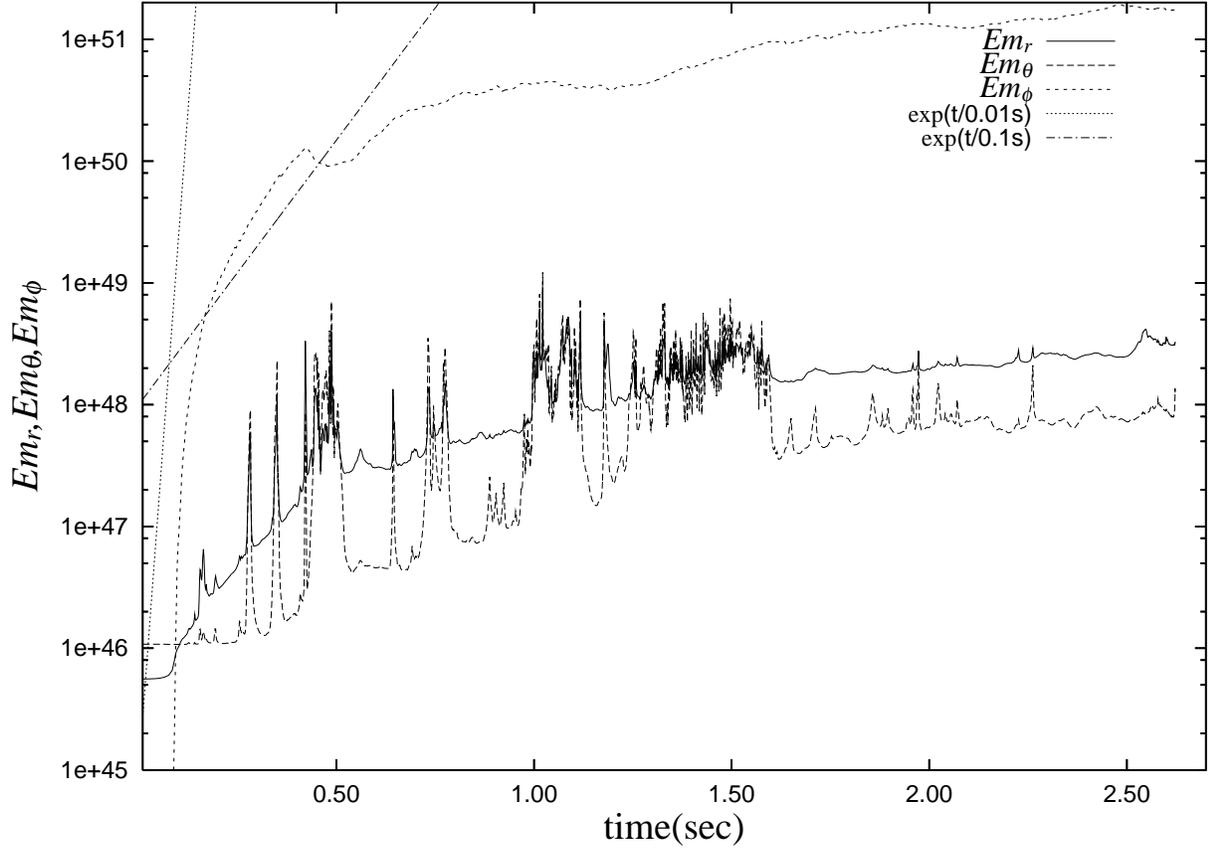}
\caption{
Time evolution of the magnetic energies
integrated over the entire computational domain, defined as 
$E_{\rm m,\,i} = \int (B_i^2/8\pi) dV$ ($i = r, \theta, \phi$).
The solid, dashed, and dotted lines represent 
$E_{\rm m,\,r}$, $E_{\rm m,\,\theta}$, and $E_{\rm m,\,\phi}$, 
respectively.} \label{fig:r10-Emag}
\end{figure}
\begin{figure}
\plottwo{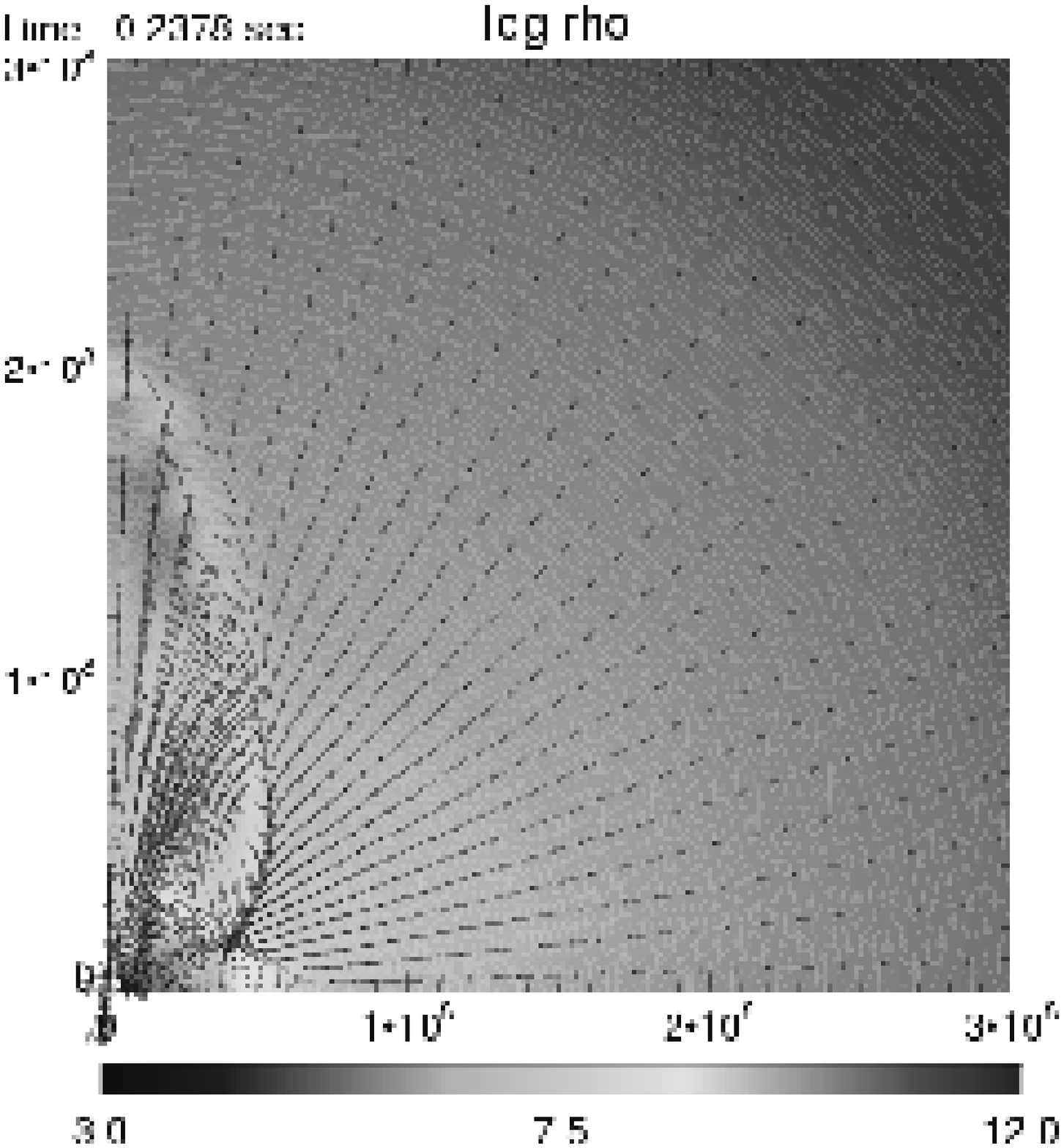}{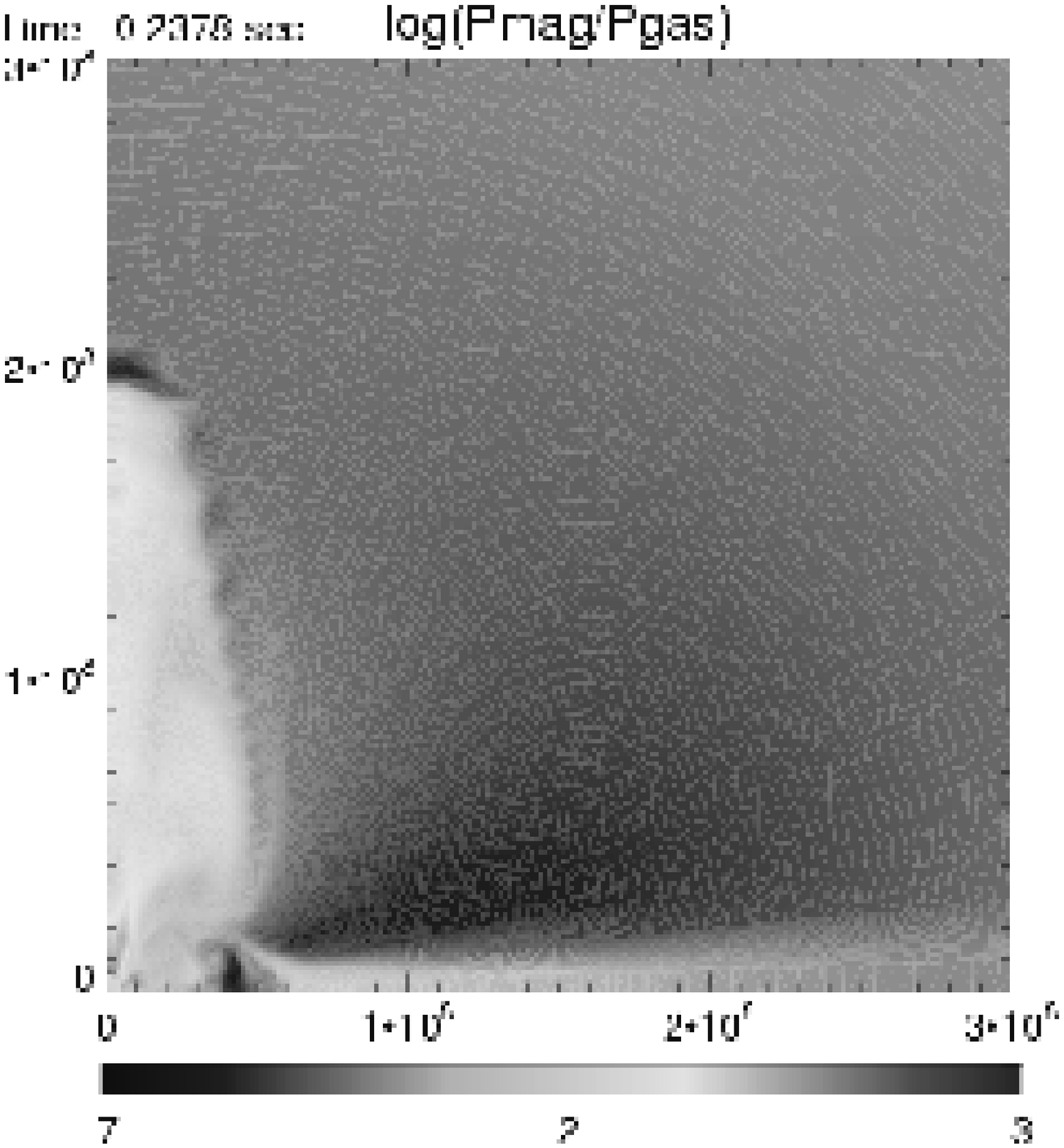}
\plottwo{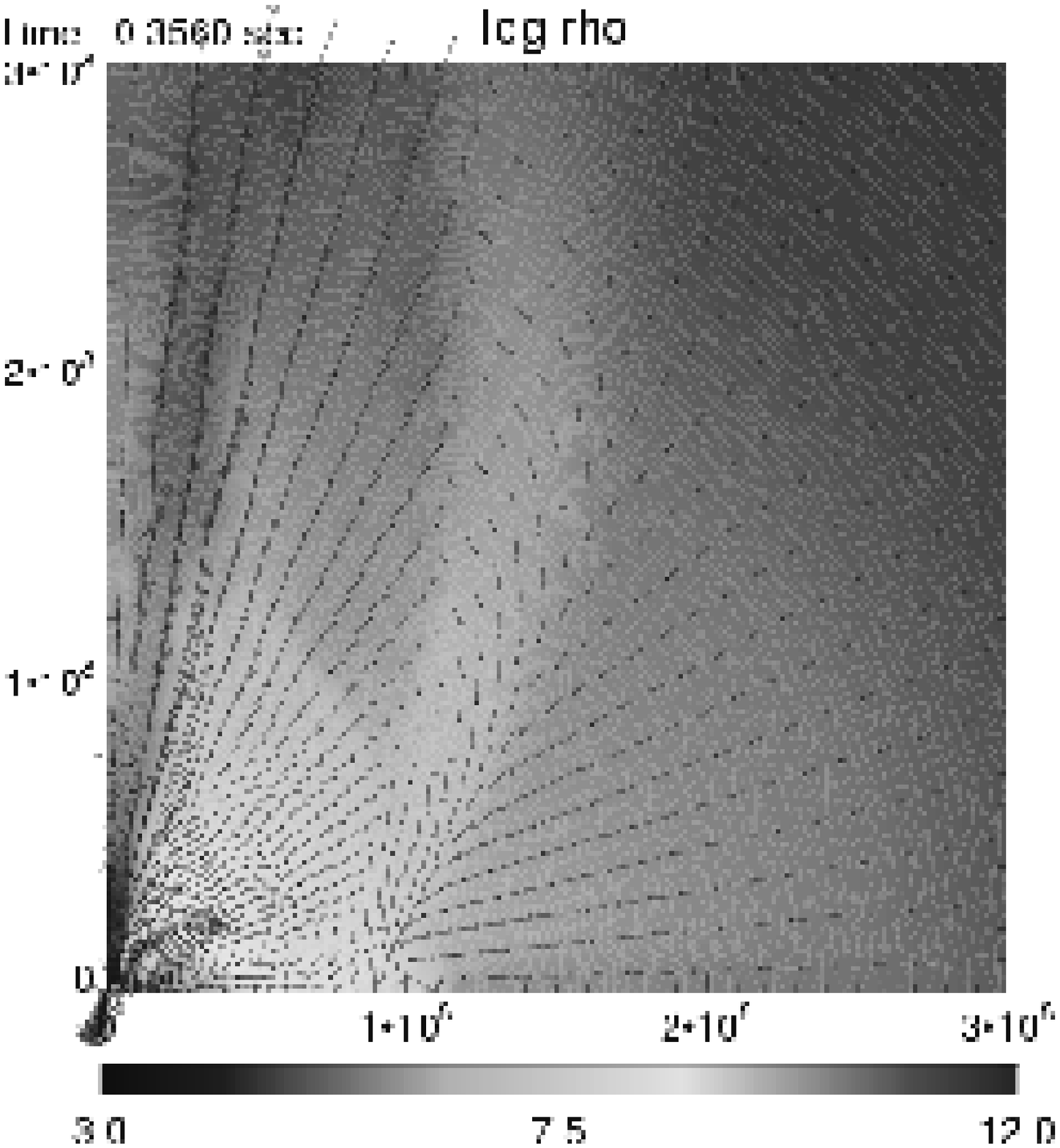}{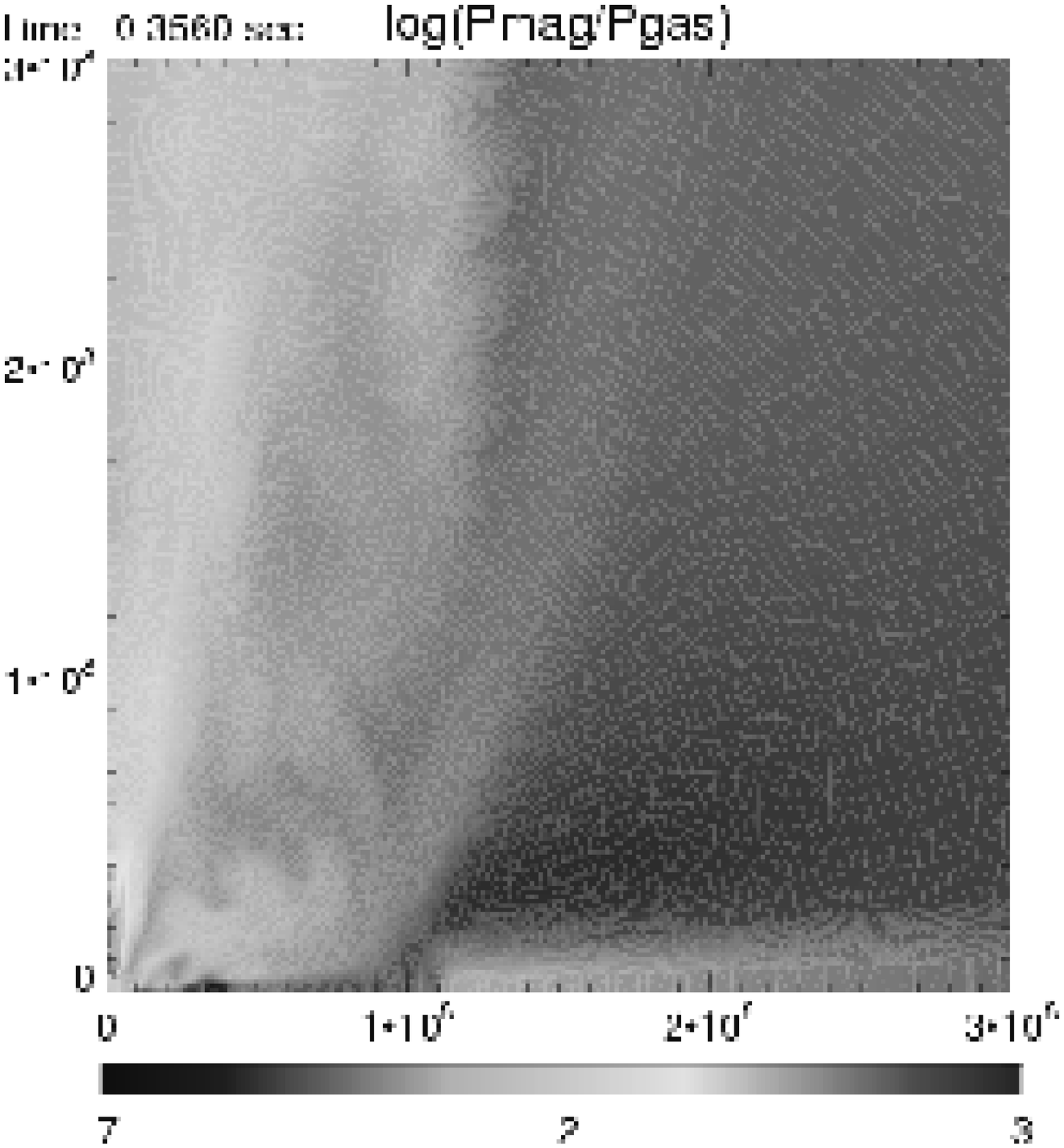}
\caption{
Contours of density (left panels) 
and the ratio of $P_{\rm mag}$ to the pressure (right panels) 
for R12 at $t = 0.2378$ and $0.3560\s$.
} \label{fig:r12-contours}
\end{figure}
\begin{figure}
 \plotone{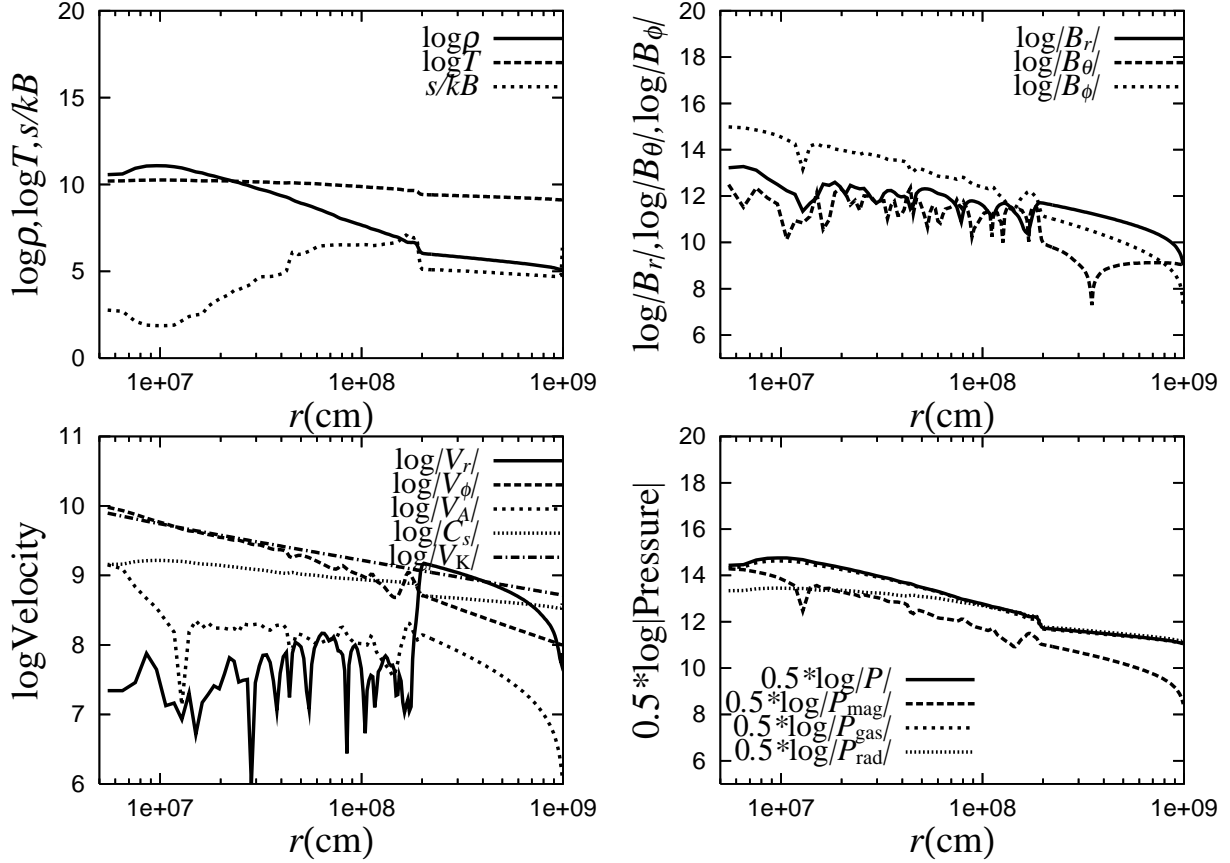}
\caption{
Physical quantities near the equatorial plane ($\theta = 88.1^\circ$)
for R10 at $1.6625\s$.
The top left panel plots the logarithmic density in $\gpccm$(solid line), 
logarithmic temperature in K (dotted line), 
and entropy per baryon in the Boltzmann constant (dashed line).
The top right panel plots logarithmic magnetic fields.
Solid, dotted, and dashed lines represents 
the radial, azimuthal, and toroidal components of the magnetic field
in units of G, respectively.
The bottom left panel plots logarithmic velocities in $\rm cm\,s^{-1}$;
the radial velocity (solid line), the angular velocity (dotted line), 
the Alfv{\'e}n speed (dashed line), the sound speed (solid-dotted line),
and the Keplerian velocity around a 2$\Ms$ black hole (solid-triple dotted line).
The bottom right panel plots logarithmic pressure in $\rm ergs\,cm^{-3}$.
} \label{fig:r10-disk}
\end{figure}
\begin{figure}
 \plotone{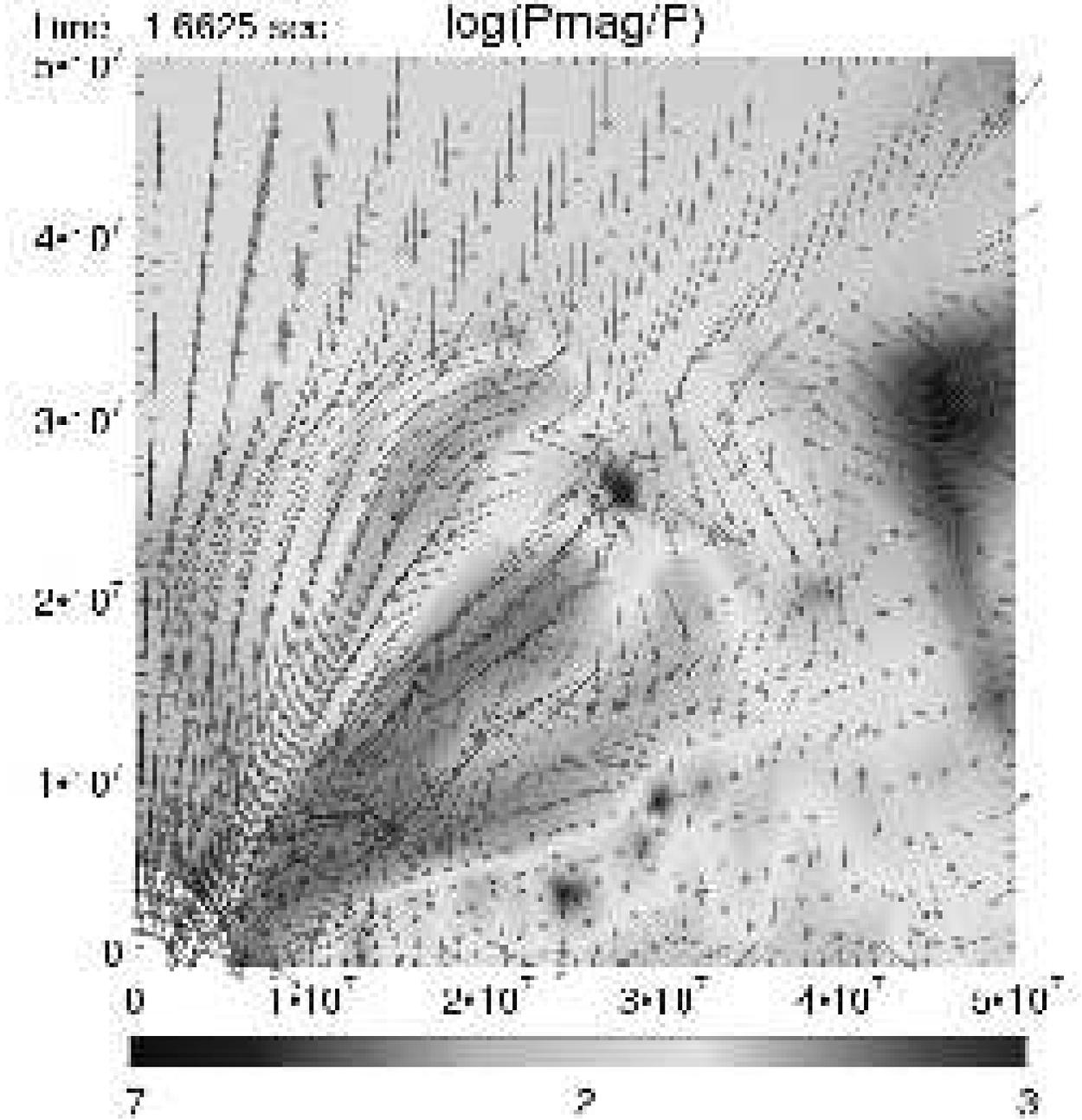}   
\caption{
Contours of the ratio of $P_{\rm mag}$ to the pressure at an
inner region (500km$\times$500km) of an accretion disk for R10
at $t = 1.6625 \s$.
} \label{fig:r10-inner-region}
\end{figure}
\begin{figure}
 \plotone{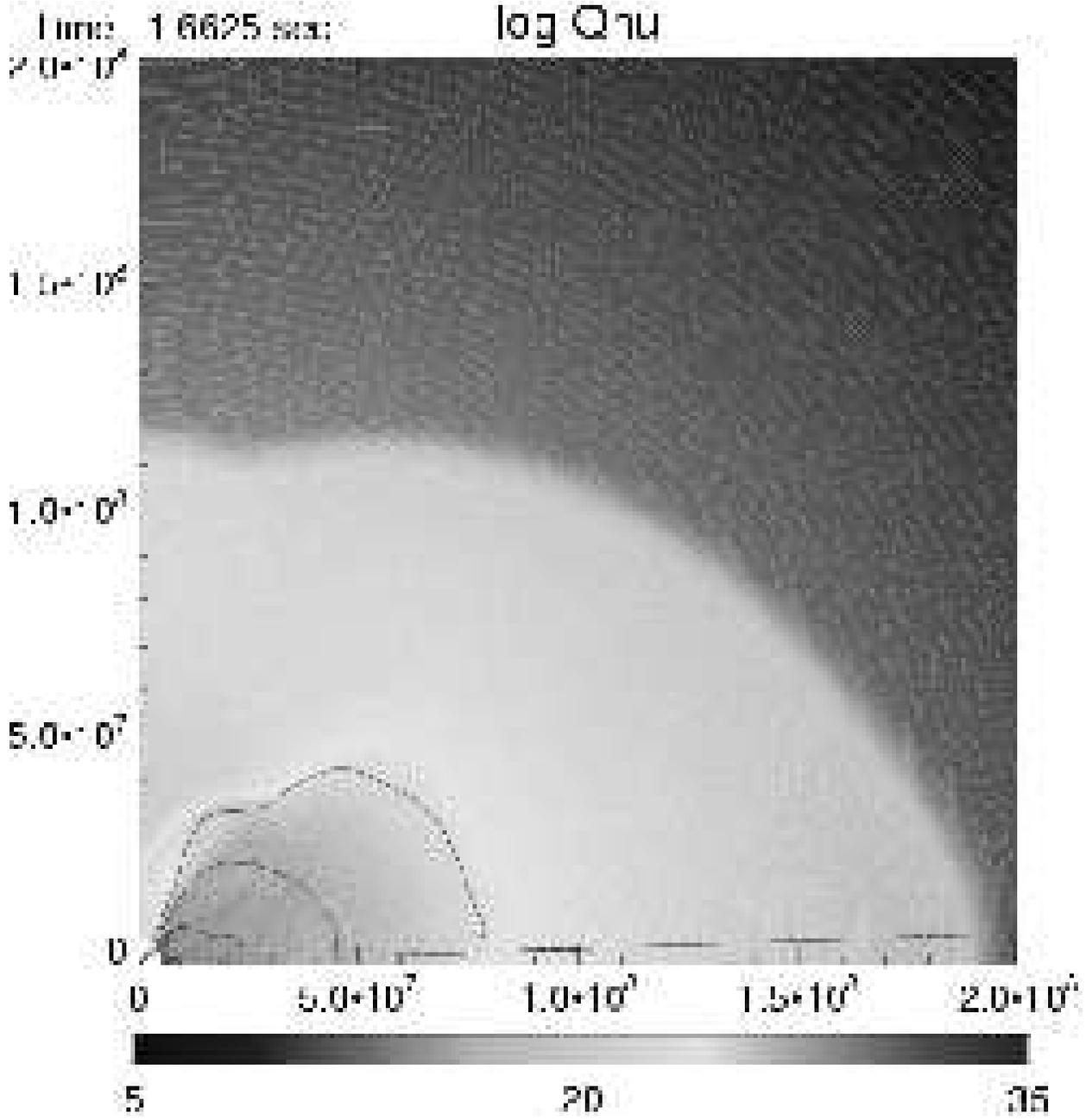}   
\caption{
Contour of logarithmic specific neutrino cooling rate 
in units of $\ergps\, cm^{-3}$ for R10 at $1.6625\s$.
Density contours of $10^8, 10^9, 10^{10}$, and $10^{11}\gpccm$
are also shown.} \label{fig:r10-q_nu}
\end{figure}
\begin{figure}
\plottwo{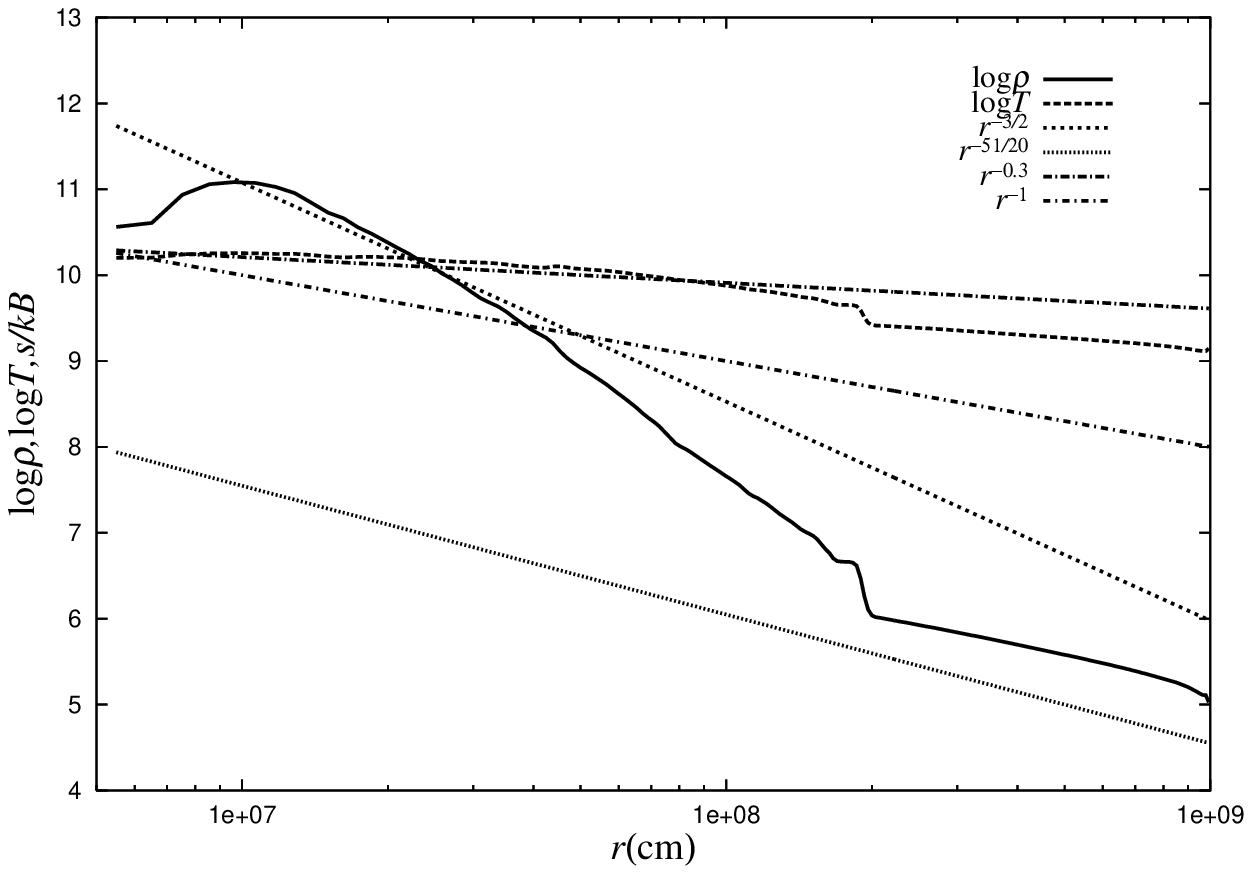}{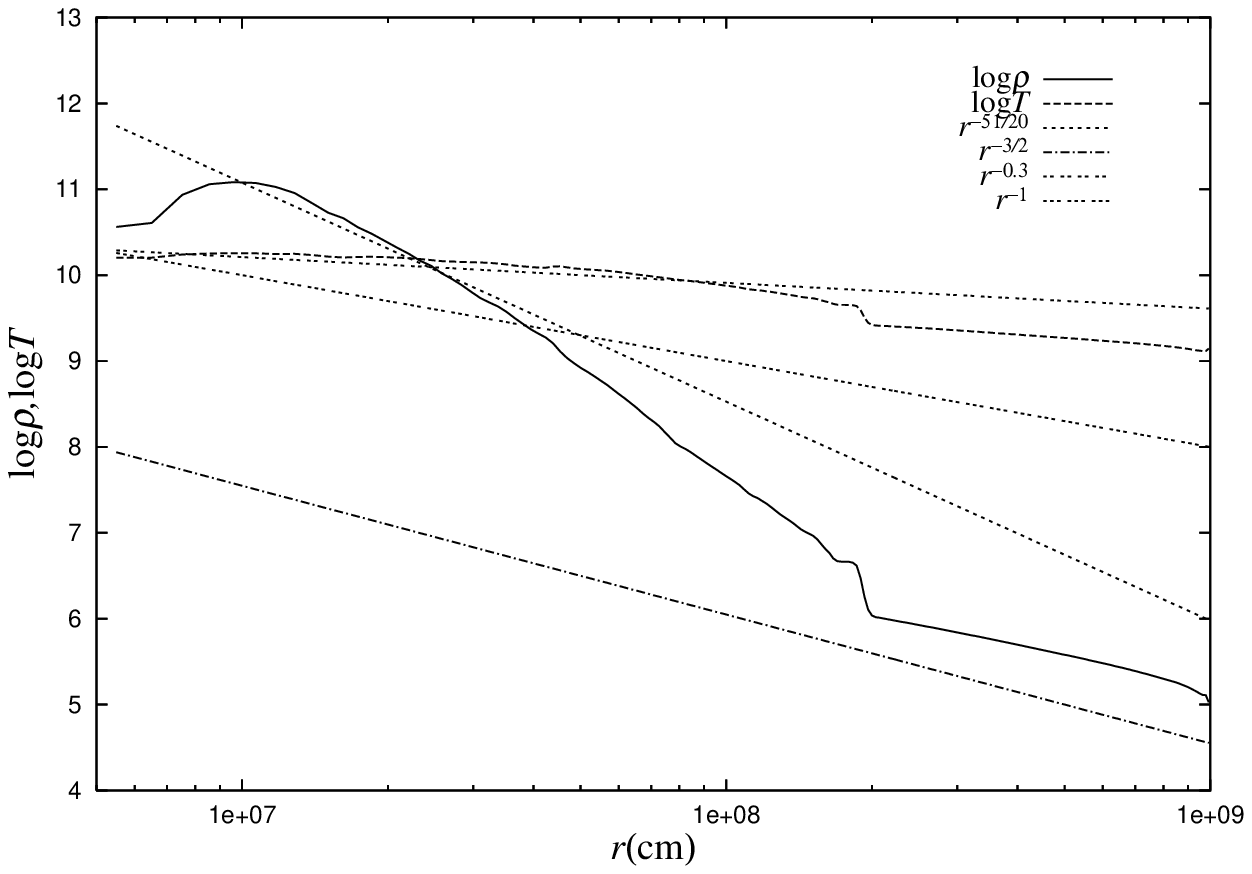}
\plottwo{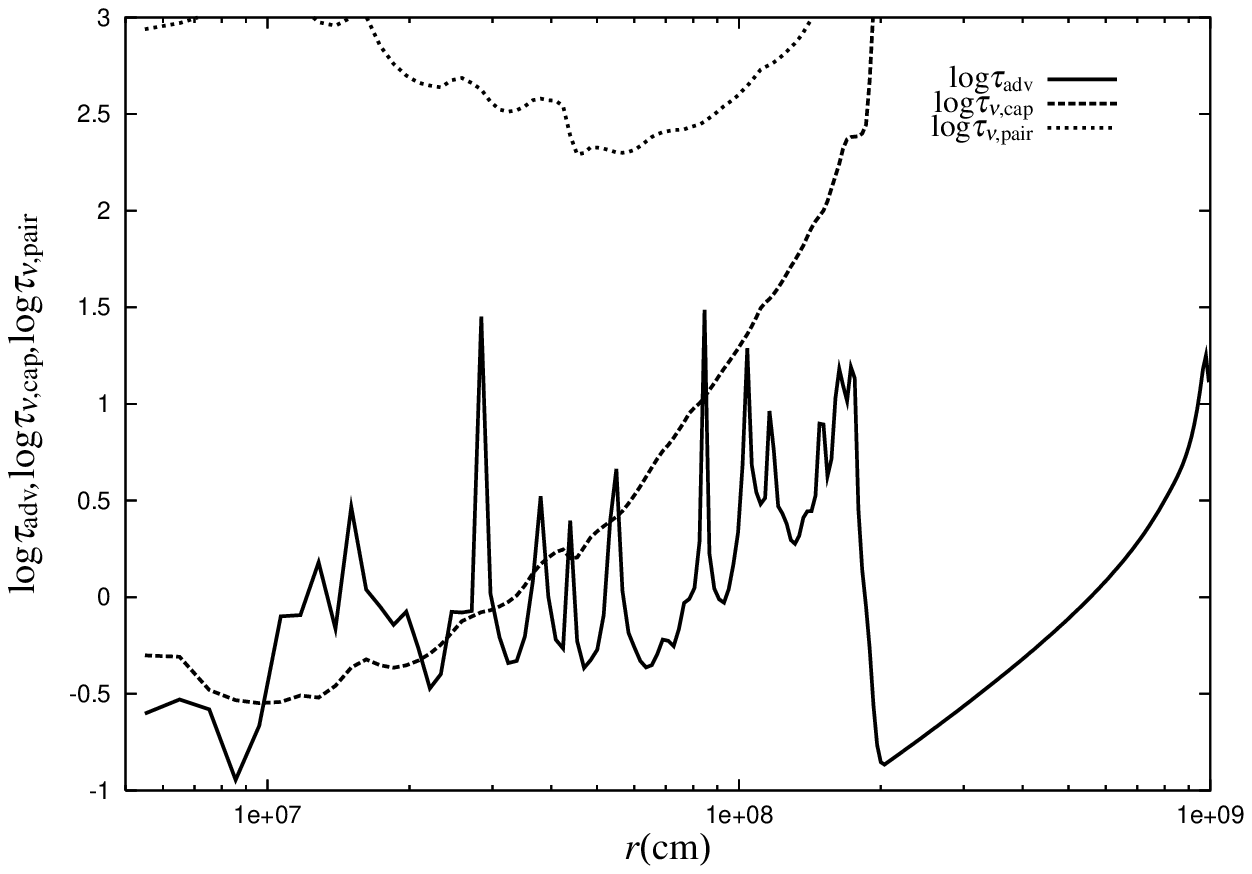}{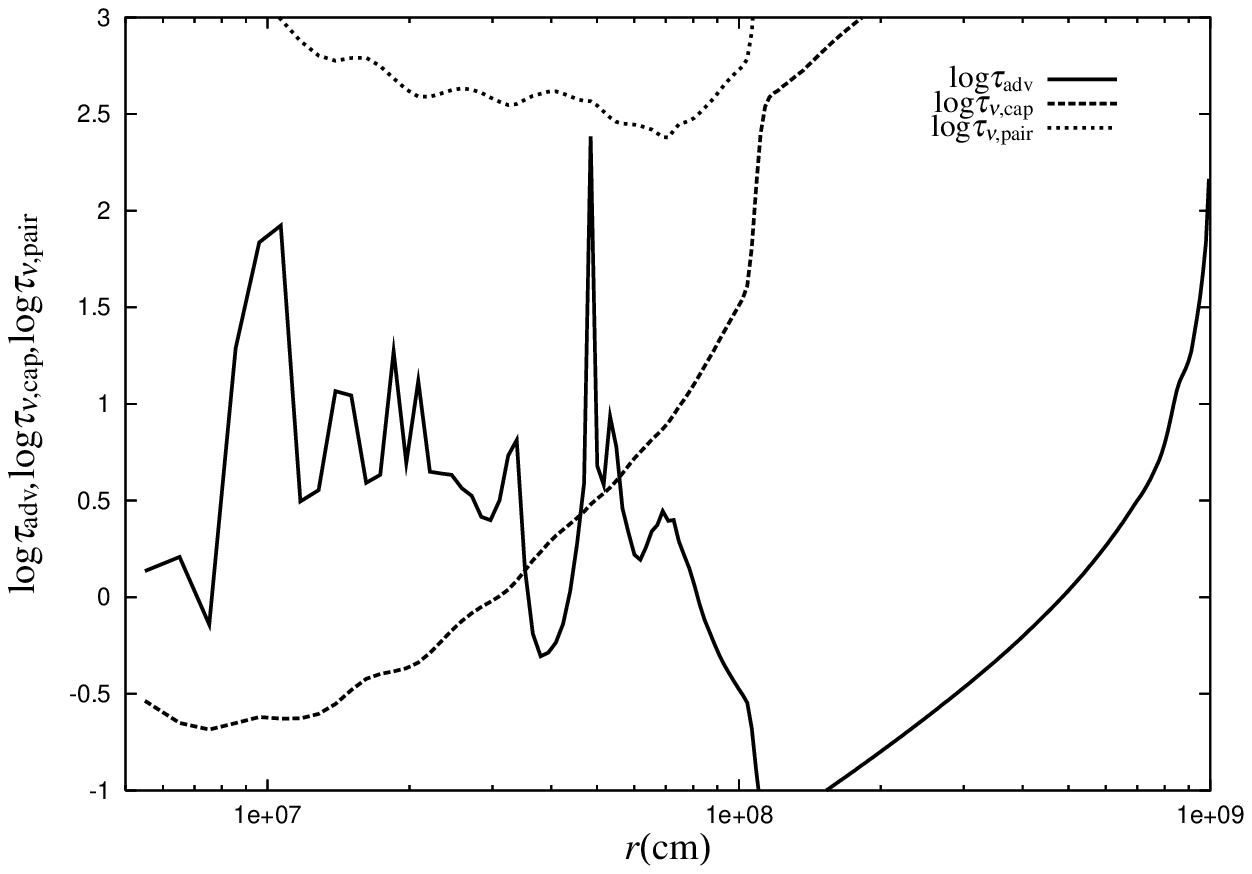}
\caption{
Radial profiles of the density and temperature in 
our models and those of ADAF and NDAF (top panels), 
and the accretion timescale and the neutrino cooling timescale
 (bottom panels) near the equatorial plane  ($\theta = 88.1^\circ$).
The top left panel plots the radial profiles of 
density (solid line) and temperature (dotted line) 
for R10 at $1.6635\s$.
Dotted and double dashed lines are scaled density and temperature
profiles of NDAF, respectively, and 
triple dashed and dot dashed lines are those of ADAF.
At an inner region, the radial profiles are well described those of NDAF.
The top right panel is same as the top left panel but 
for R8 at 4.0000 s.
The bottom left panel plots accretion timescale (solid line) and
neutrino cooling timescale through pair capture (dashed line) and
pair annihilation (dotted line) for R10 at $1.6635\s$.
The top right panel is same as the bottom left panel but 
for R8 at $4.0000 \s$.
} \label{fig:r10-disk-pl}
\end{figure}
\begin{figure}
 \plotone{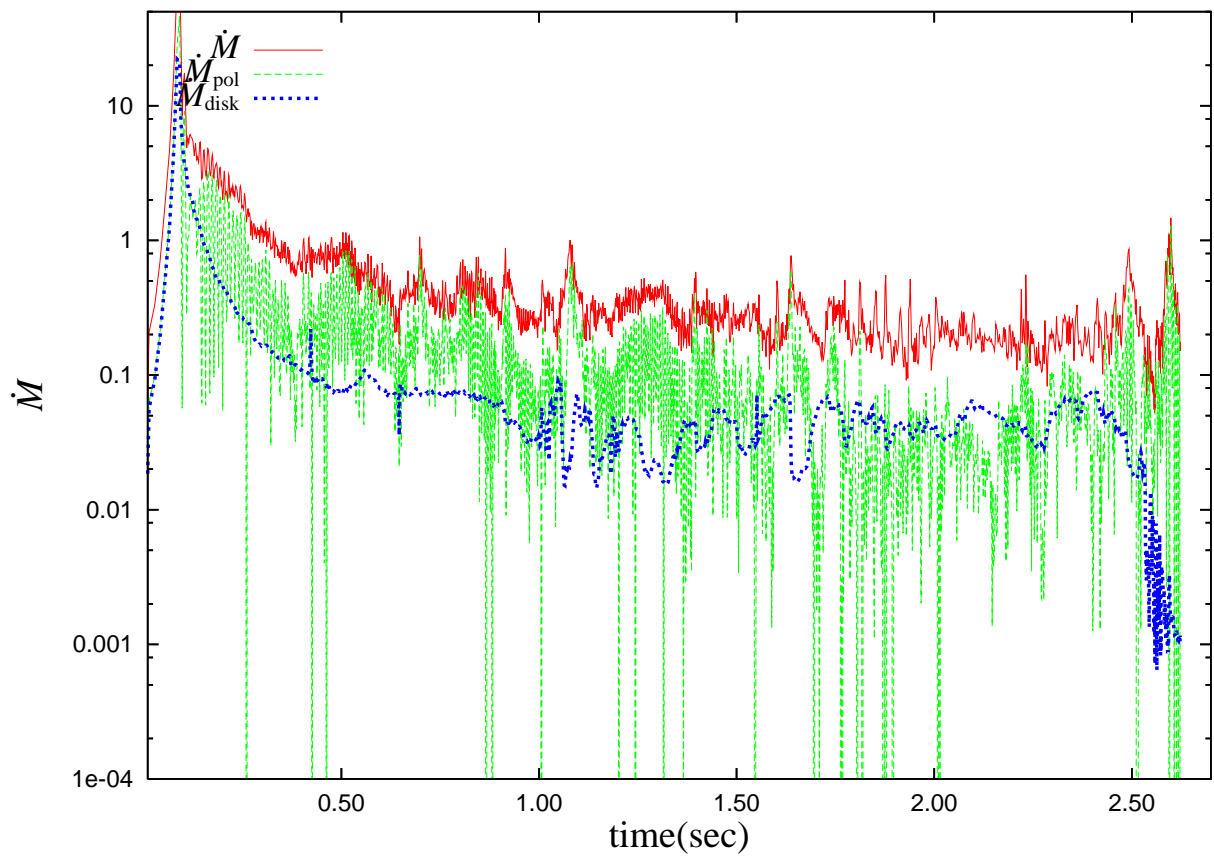}
\caption{
Time evolution of the accretion rates.
The solid, dashed, and dotted lines represent 
the rate through the inner boundary at 50km, $\dot{M}$, 
the rate through parts of the boundary ($\theta \le 20^\circ$), 
$\dot{M}_{\rm pol}$ 
the rate through parts of the boundary ($\theta \ge 50^\circ$), 
$\dot{M}_{\rm disk}$, respectively.
All rates are in units of $\DMs$.}
\label{fig:r10-Mdot}
\end{figure}
\begin{figure}
 \plotone{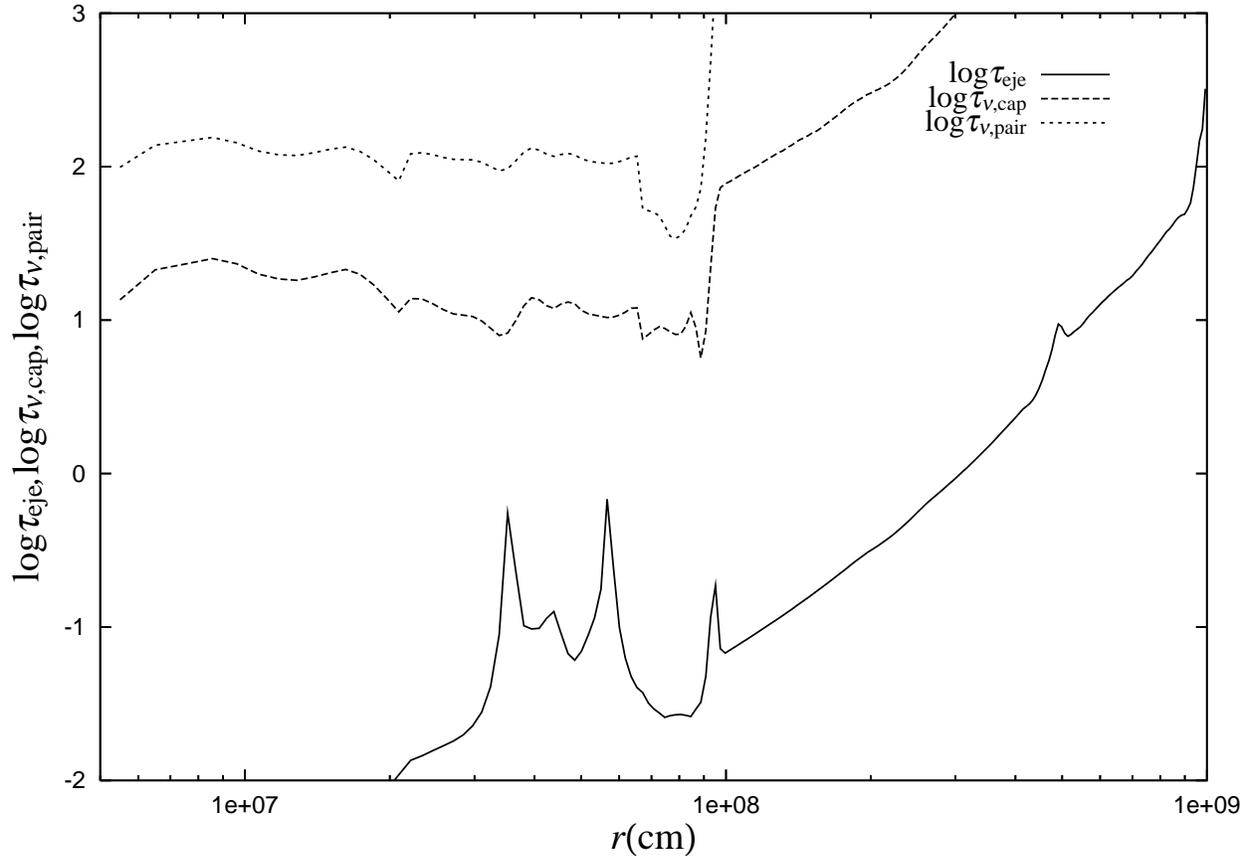}
\caption{
Timescale inside a jets for R12 at $0.1997\s$.
Solid, dashed, and dotted lines are 
ejection timescale, neutrino cooling timescale through pair capture
and that through pair annihilation.
} \label{fig:r12-jet-tscale}
\end{figure}
\begin{figure}
\plottwo{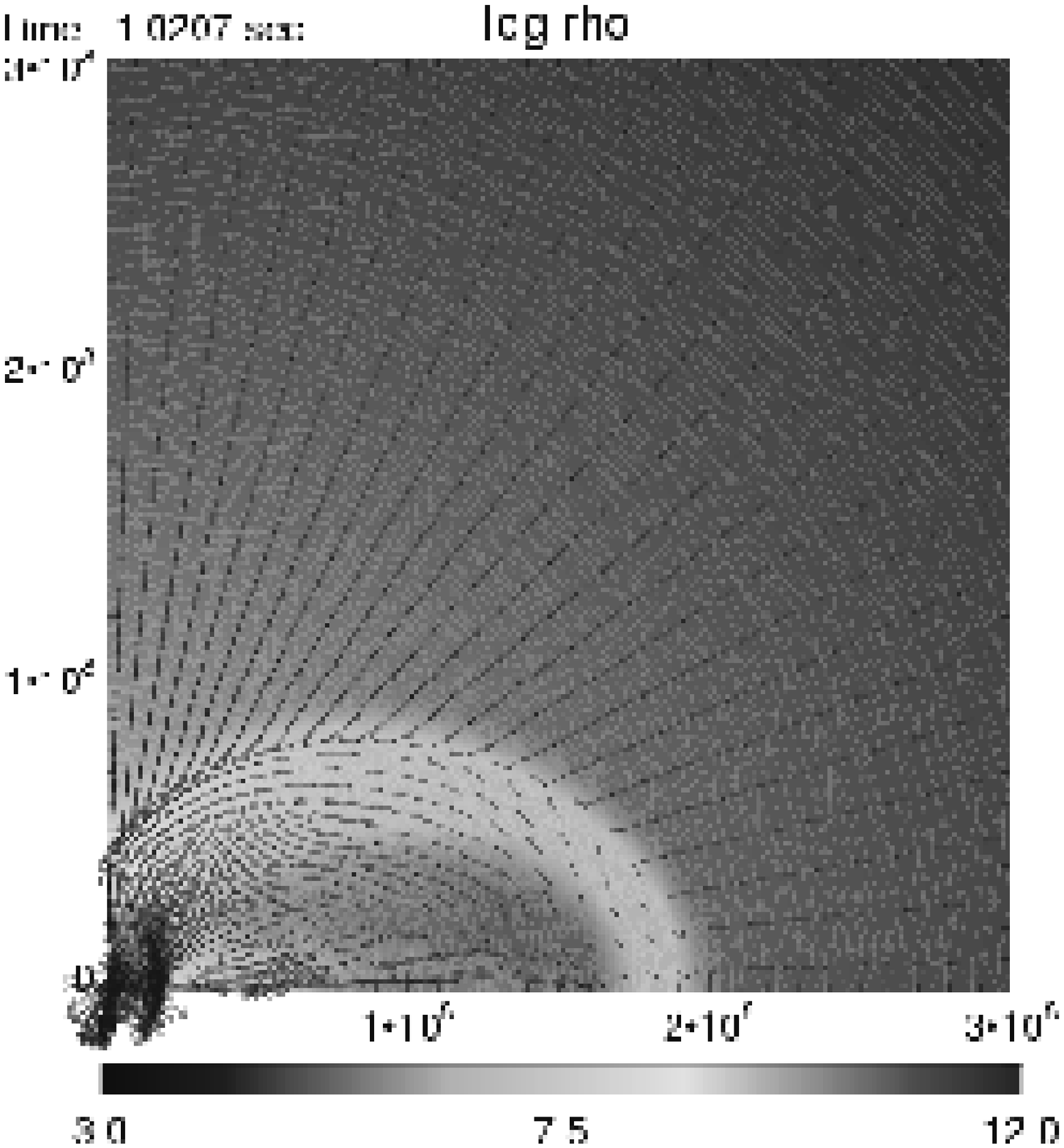}{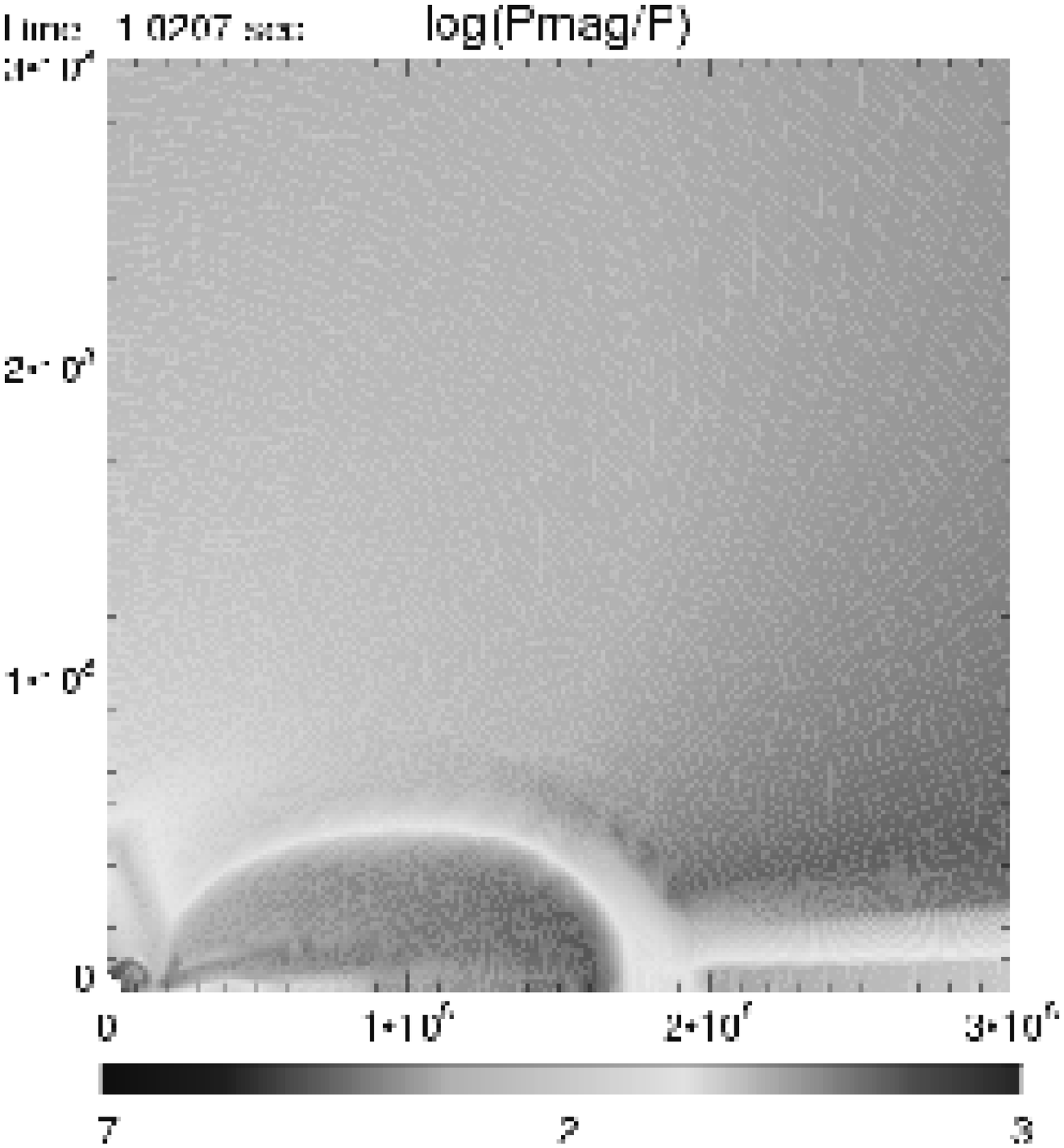}
\plottwo{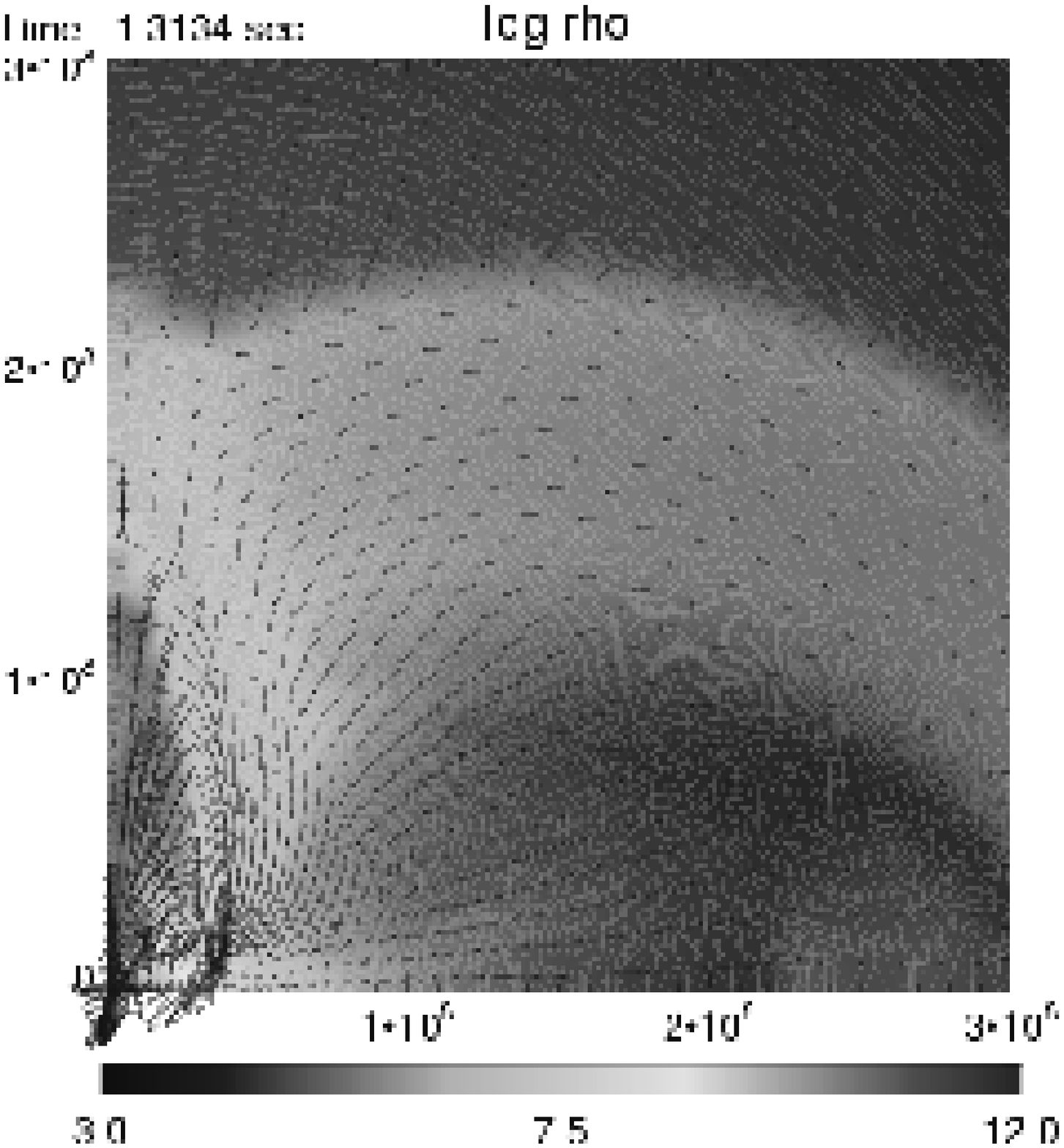}{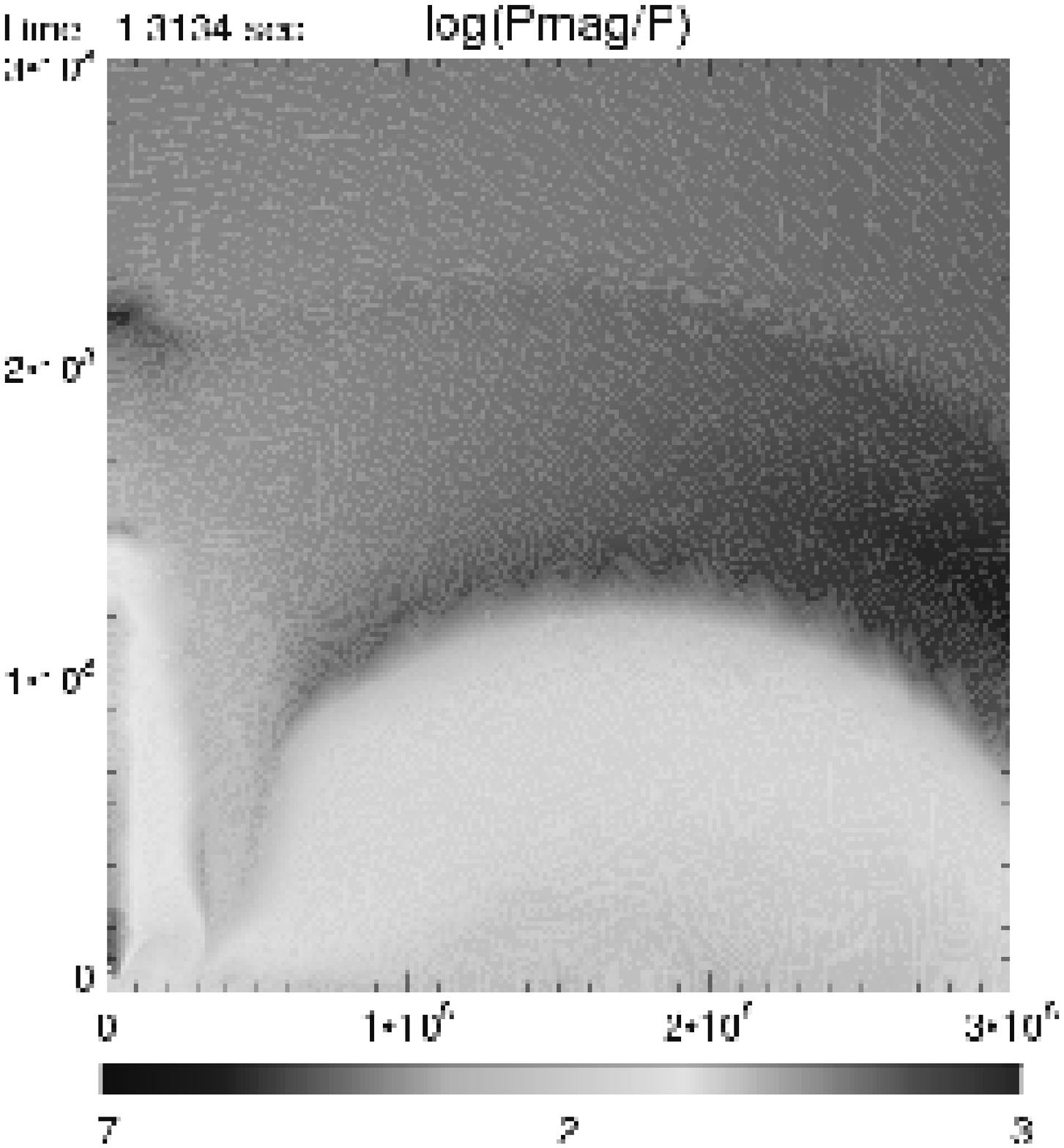}
\caption{
Contours of density (left panels) 
and the ratio of $P_{\rm mag}$ to the pressure (right panels) 
for S10 at $t = 1.0207$ and $1.3134 \s$.
} \label{fig:s10-contours}
\end{figure}

Now we move on to models with slowly rotating core, or S8, S10, and S12.
As we can see in Figure \ref{fig:init-ang-momentum}, 
the initial angular momentum for these models is smaller than 
$10^{17}\rm cm^2 \rm \, s^{-1}$ except in the outer part of the computational
domain. Most of the inner region of the star therefore collapse to a
black hole. Material with high angular momentum is expected to form
a disk-like structure.
Figure \ref{fig:s10-contours} shows contours of density (left panels)
and those of the ratio of $P_{\rm mag}$ to the pressure (right panels) 
for S10 at $t = 1.0207$ and $1.3134 \s$.
The core for S10 rotates slower than that for R10.
The star collapses with nearly spherical configuration for a longer time.
When the rapidly rotating gas falls near the black hole $(t \sim 0.45 \s)$,
the magnetic field has been amplified due to field compression and
wrapping initially and MRI later.
Eventually, a magnetically driven jet is found to be 
generated from a central region near the black hole 
(bottom panels in Figure \ref{fig:s10-contours}), 
although the core rotates slowly.
We find that the jet is less collimated than that in R10 
(bottom left panels in Figures \ref{fig:r10-contours} and 
\ref{fig:s10-contours}).

\begin{figure}
 \plotone{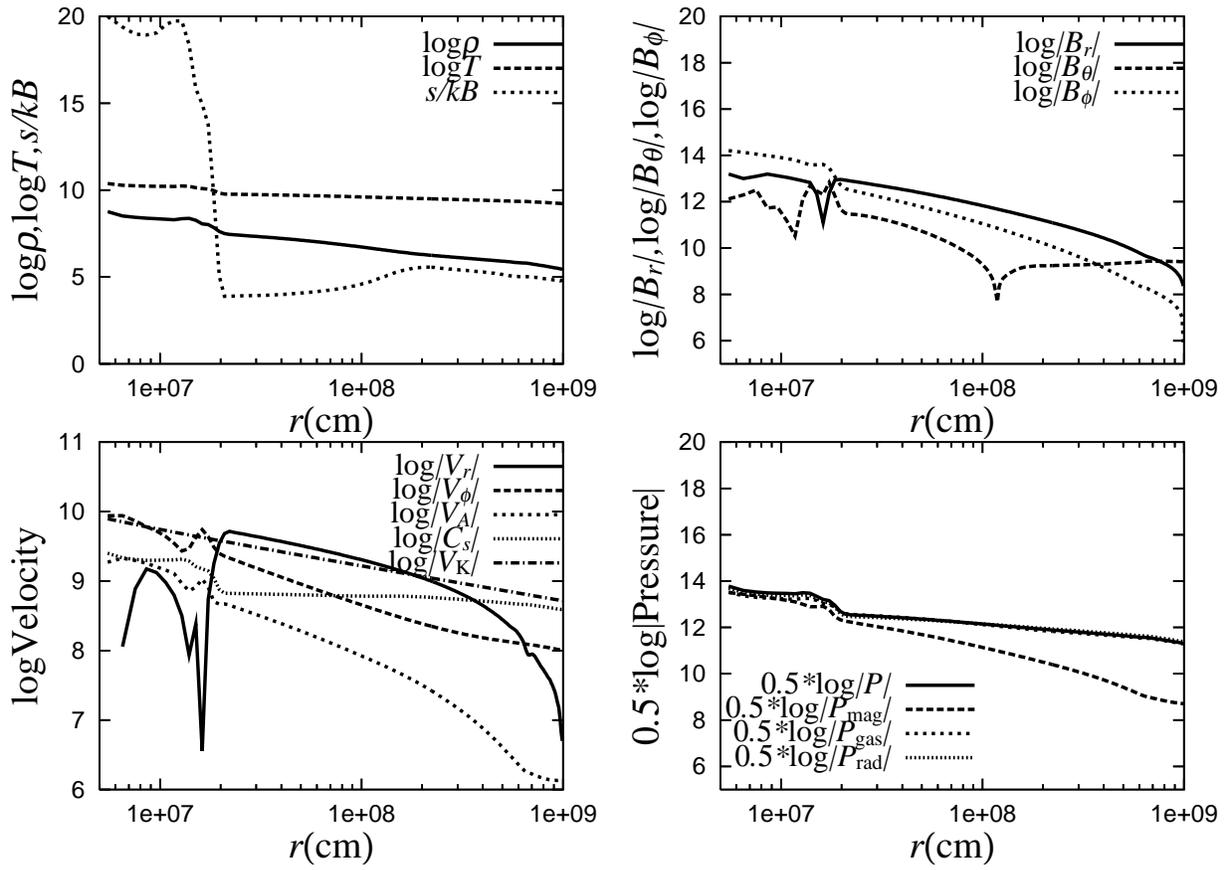}
\caption{
Same figure as Figure \ref{fig:r10-disk} but for S10 at $0.5676\s$.
} \label{fig:s10-disk}
\end{figure}

A disk like structure is formed as in R10 but
magnetically driven winds arise near the equatorial plane 
(top left panel in Figure \ref{fig:s10-contours}), 
which disappear in R10.
Figure \ref{fig:s10-disk} shows physical quantities 
near the equatorial plane ($\theta = 88.1^\circ$)
for S10 at $t = 0.5676 \s$.
When gas infall to the remnant with high angular momentum, 
the magnetic field has been amplified so large that 
the magnetic pressure is comparable to the gas pressure 
near the central remnant.
Consequently, the winds are driven via magnetic and gas pressure
for S10, in contrast to R10.
These magnetically driven winds have high velocity 
($\sim 0.1c$; bottom left panel in Figure \ref{fig:s10-disk}) 
and relatively high entropy of $s \sim 20$ 
(top left panel in Figure \ref{fig:s10-disk})
where $s$ is the entropy per baryon in units of
the Boltzmann constant.
Distributions of the density, temperature and pressure 
cannot be described with those of simple power laws in $r$, 
in contrast to R10, due to the winds.

\begin{figure}
\plottwo{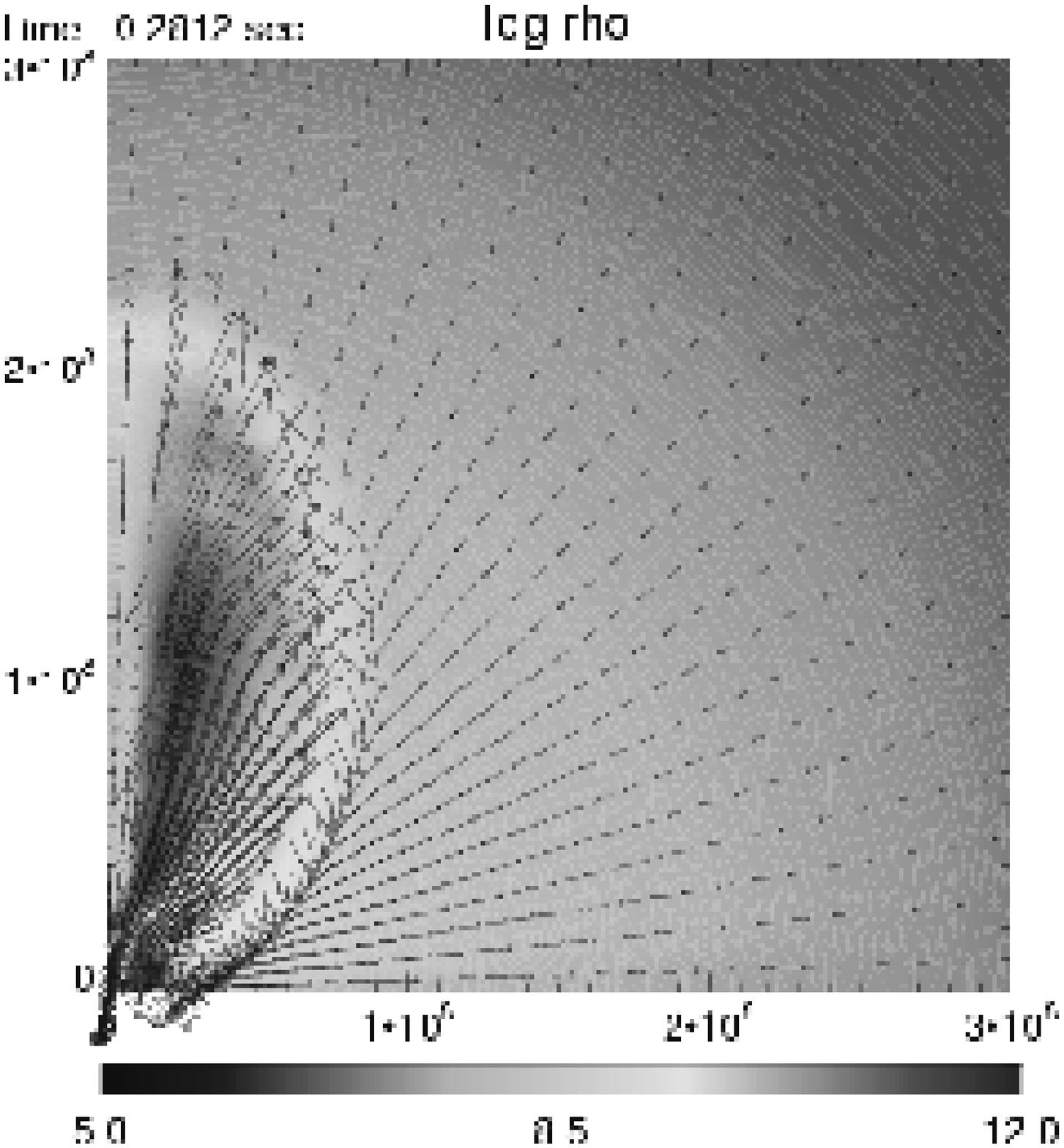}{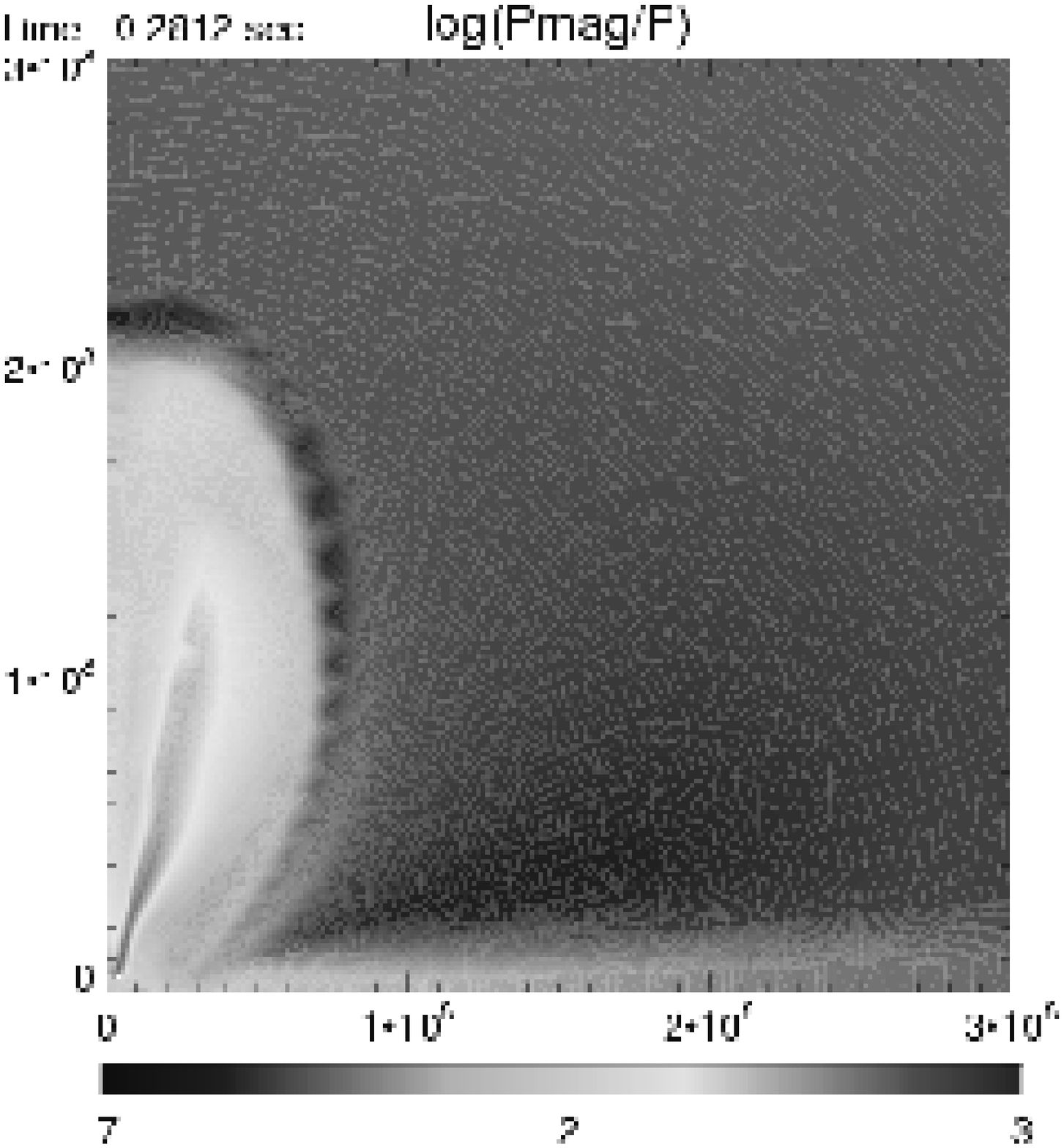}
\caption{
Contours of density (left panel) 
and the ratio of $P_{\rm mag}$ to the pressure (right panel) 
for S12 at $t = 0.2812 \s$.
} \label{fig:s12-contours}
\end{figure}

As in cases with rapidly rotating core, 
the toroidal magnetic field is more rapidly amplified
in models with a larger initial magnetic field.
We find that 
S12 also launches a jet from a central region near the black hole 
along the rotational axis.
Figure \ref{fig:s12-contours} shows contours of density (left panel)
and those of the ratio of $P_{\rm mag}$ to the pressure (right panel) 
for S12 at $t = 0.2812 \s$.
The jet is collimated in a wider region compared with that in R12
(bottom panels in Figure \ref{fig:s12-contours}) because of slow core
rotation in S12.
It is noted that the jet has density lower than that in R12.

For S8, as in R8 and R10, 
a disk-like structure is formed inside a shock surface located at 800km 
on the equator at the end of the computation, 
and the disk is well described with a quasi-steady NDAF
rotating with a Keplerian profile.
The magnetic pressure, whose toroidal component attains to $10^{15}$ G, 
is lower than the gas pressure elsewhere.
Magneto-centrifugal winds are not appeared in S8, in contrast to S10,
and jet cannot be produced in S8, as in R8.

In brief, 
the collapsing star can originate well collimated jets
if the star has sufficiently large magnetic field and 
rapidly rotating core.
On the other hands, 
when the star has not rapidly rotating core but 
envelope with sufficiently high angular momentum,
the star can drive less collimated jets.
The jet could be generated from collapsing stars 
with non-rotating core and 
sufficiently high angular momentum envelope 
and large magnetic field.

\subsection{Time Evolution of Collapsing Star} \label{sec:evolution}

We present the time evolution of various quantities, such as 
the magnetic energy, the mass accretion rate, and the neutrino luminosity
of our models to compare properties of models each other. 

\begin{figure}
 \plotone{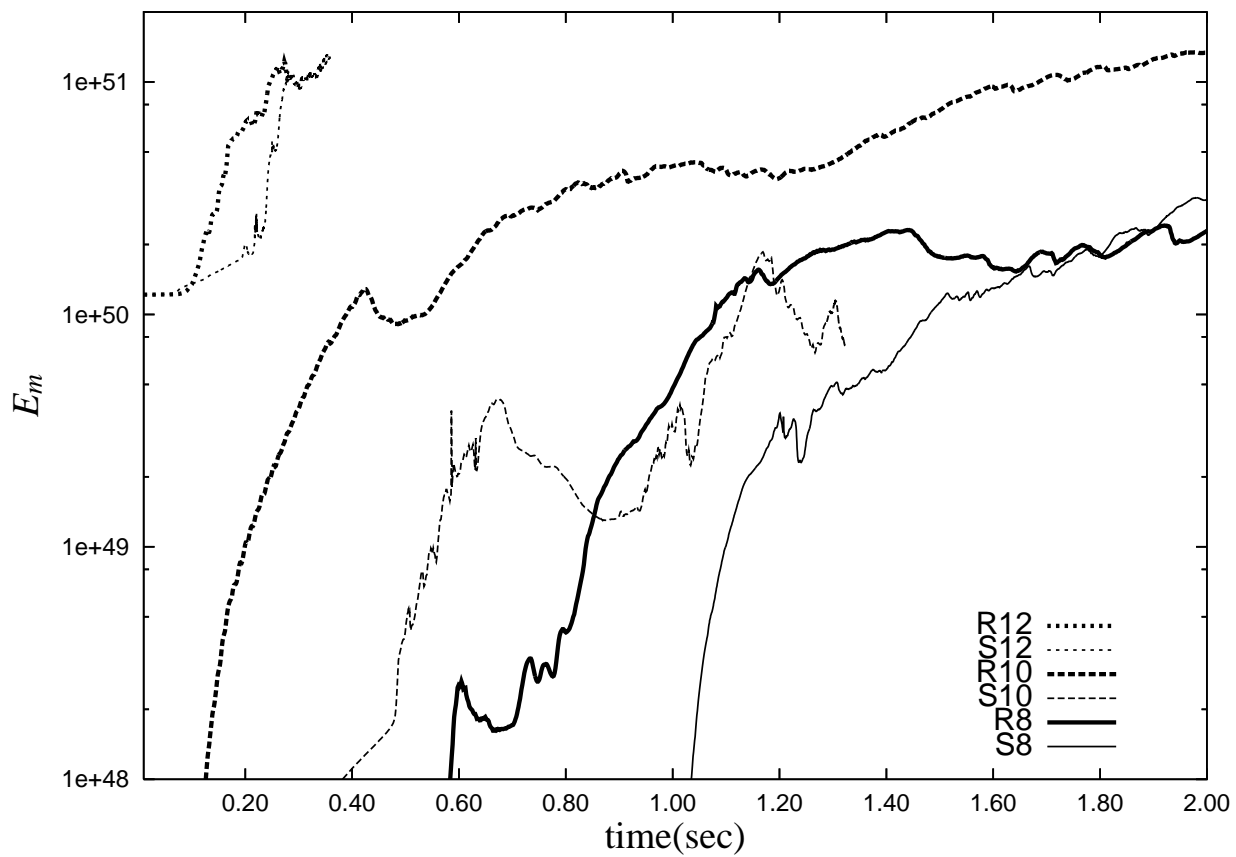}
\caption{
Time evolution of 
the magnetic energy integrated over the computational domain, 
$E_{\rm m} = \int (B^2/8\pi) dV$, for all models.
The solid, thin-solid, dashed, thin-dashed, dotted, and thin-dotted
lines represent the magnetic energy for R8, S8, R10, S10, R12, and S12, 
respectively.
} \label{fig:Emag}
\end{figure}

Figure \ref{fig:Emag} shows the time evolution of 
the magnetic energy integrated over the computational domain, 
$E_{m} = \int (B^2/8\pi) dV$, for all models.
We note that the toroidal field dominates over the other components of
the field, as in R10 (Figure \ref{fig:r10-Emag}).
The fields are amplified earlier as the initial magnetic field increases
and the core rotation becomes faster.
This is because the magnetic field is mainly amplified by the wrapping
of the field. 
Rapid differential rotation is therefore required from the amplification.

\begin{figure}
\epsscale{.8}
 \plotone{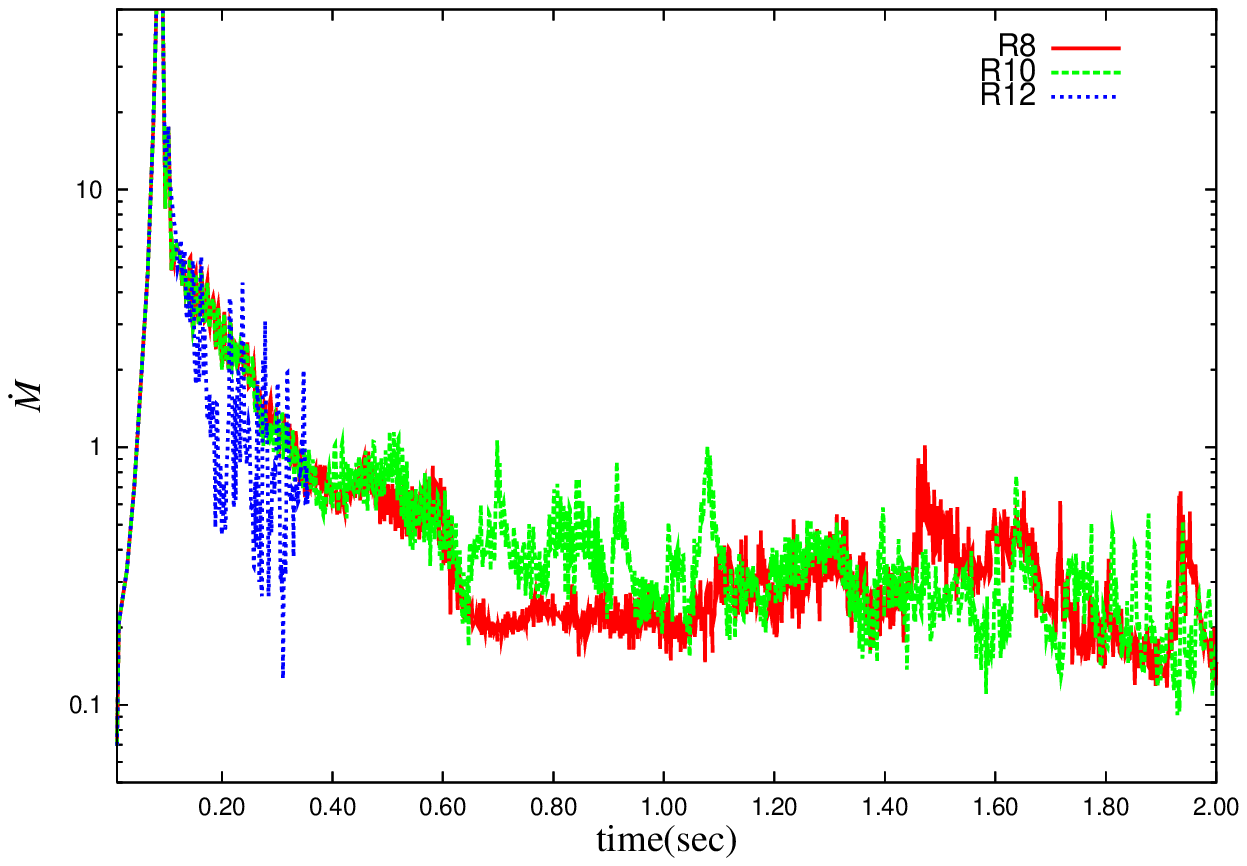}
 \plotone{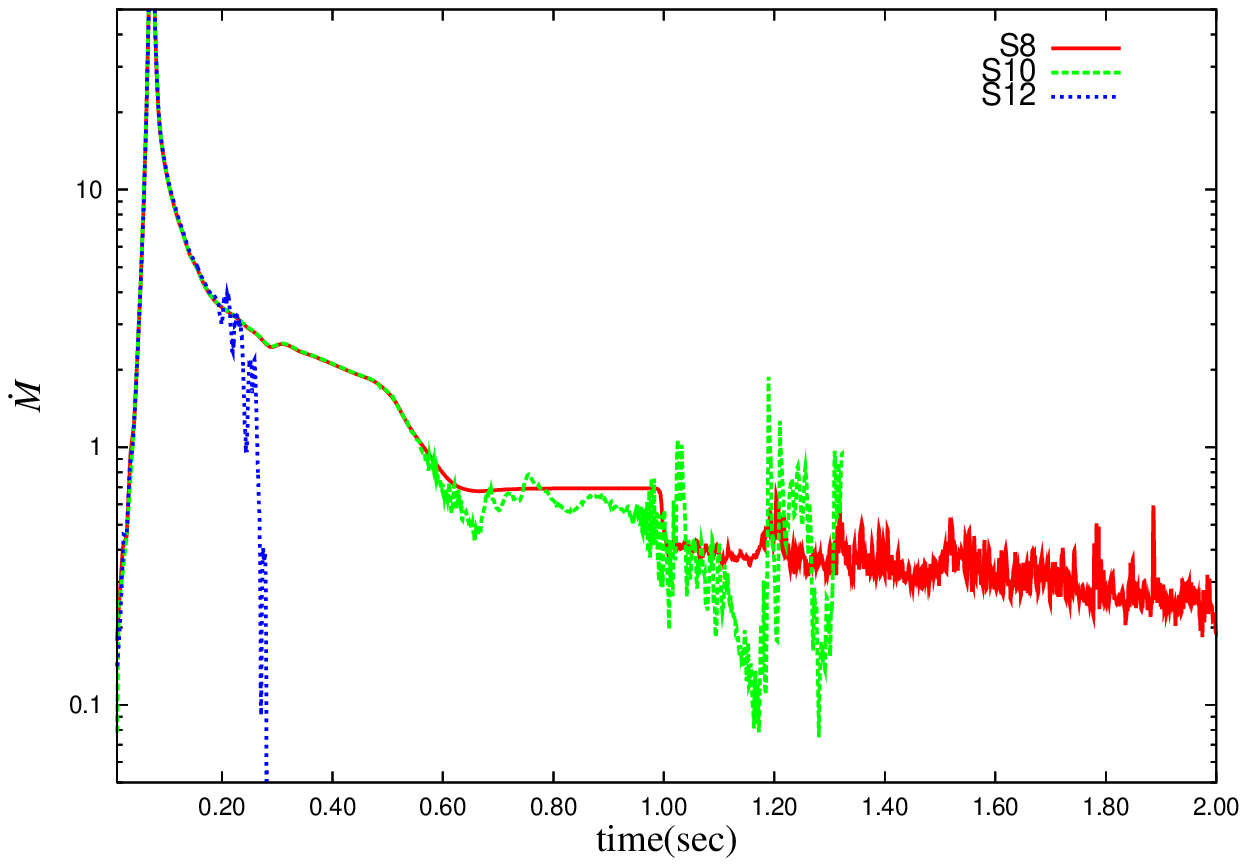}
\epsscale{1}
\caption{
Time evolution of the accretion rate 
through the inner boundary at 50km, $\dot{M}$, for all models.
The solid, dashed, and dotted lines represent 
the rates for R8, R10, and, R12 (left panel), respectively, 
and the rates for S8, S10, and, S12 (right panel), respectively.
} \label{fig:Mdot}
\end{figure}
\begin{figure}
 \plotone{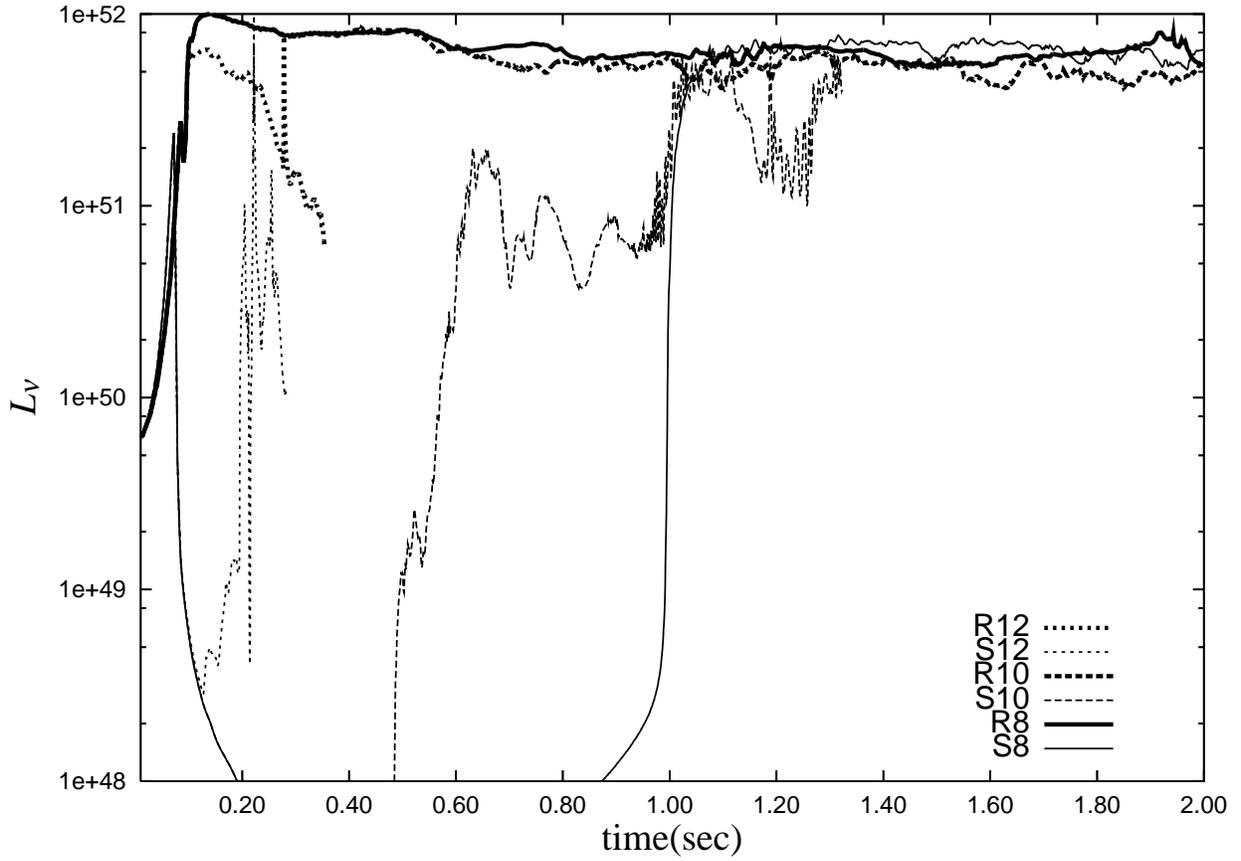}
\caption{
Time evolution of 
the neutrino luminosity integrated over the computational domain, for all models.
The solid, thin-solid, dashed, thin-dashed, dotted, and thin-dotted
lines represent the neutrino flux for R8, S8, R10, S10, R12, and S12, 
respectively.
} \label{fig:Lnu}
\end{figure}

Figure \ref{fig:Mdot} shows the time evolution of $\dot{M}$
for rapidly rotating models, or R models (upper panel) and 
slowly rotating models, or S models (bottom panel). 
At the beginning of the simulations ($t < 0.1 \s$), 
$\dot{M}$ roughly are comparable for all models
because the collapse is almost spherical and 
effects of rotation and magnetic field can be neglected in this phase.
After the infall of higher angular momentum material, 
$\dot{M}$ rapidly drops due to the centrifugal barrier.
This is because $\dot{M}$ in R models are smaller than 
those of S models during an initial phase ($t \le 0.5 \s$).
Once a disk-like structure is formed (for R8, R10, S8, and S10), 
$\dot{M}$ stays 0.2-0.3 $\DMs$ with fast fluctuations.
We note that $\dot{M}_{\rm disk}$ is much lower than $\dot{M}$ by
about an order of magnitude for all models, as in R10 
(Figure \ref{fig:r10-Mdot}).


Figure \ref{fig:Lnu} shows the time evolution of 
the neutrino flux integrated over the computational domain,
or the neutrino luminosity, $L_{\nu}$, for all models.
Neutrino is initially emitted from a quasi-spherically collapsing dense core 
near the black hole at the maximum rate of $\sim 10^{52} \ergps$
around $\sim 0.1\s$.
For R8 and R10, material with high angular momentum reaches to the hole 
before the infall of the entire dense core to the hole.
Neutrino continues to be emitted from a disk formed near the equator 
(Figure \ref{fig:r10-q_nu} for R10)
so that $L_{\nu}$ stays roughly constant at the rate of 
5-6 $\times 10^{51} \ergps$.
Note that the neutrino luminosity is much smaller than 
that for SN $\sim$ 5-6 $\times 10^{52} \ergps$ 
at $0.1\s$ after core bounce~\citep{lieb01}
because in SN, neutrino is efficiently emitted from 
a region near a proto neutron star.
We note that 
the neutrino luminosity slightly increases if we adopt 
much smaller inner boundary, 
as we shall see later (\S \ref{sec:inner-boundary}).
On the other hand, for S8 and S10, 
high angular momentum material falls to the hole 
after the accretion of the entire dense core.
$L_{\nu}$ therefore drops sharply around $0.1 \s$ and 
raises rapidly when a rotationally supported disk is formed 
near the equator ($t \sim 1.0 \s$ and $0.6 \s$ for S8 and S10, respectively).

\section{Discussion} \label{sec:discussion}

\subsection{Numerical resolution of our simulations} \label{sec:resolution}

If the fastest growing mode of MRI is unresolved, 
saturation of the MRI occurs later and at a lower amplitude of the 
magnetic energy~\citep{sp01}.
Hence, we check whether our numerical grid can resolve the fastest growing
mode of MRI, whose wavelength is given by 
$\lambda_c = 2\pi v_{\rm A}/\sqrt{3}\Omega$ where $v_{\rm A}$ is 
the Alfv{\'e}n velocity.
We need to verify the relation, $\Delta r < \lambda_c$, which leads to
\begin{equation}
 \frac{\Delta r}{r} < \frac{2\pi v_{\rm A}}{\sqrt{3}v_\phi},
 \label{eq:mri_resol}
\end{equation}
where $\Delta r$ is the grid size.
At the beginning of simulations, 
the relation (\ref{eq:mri_resol}) cannot hold
because of small $v_{\rm A}$ compared with $v_\phi$.
However, the core is stable against MRI in this phase 
because $\Omega$ is nearly constant 
(see the initial distribution (eq. \ref{eq:omega0})).

After the formation of a disk-like structure, 
the disk becomes to be unstable to MRI.
The growth time scale of MRI, $\tau_{\rm MRI}$, 
is comparable to that of the wrapping of the field
$\tau_{\rm wrap}$ ($\sim 0.1 \s$).
However, the relation (\ref{eq:mri_resol}) cannot hold.
The magnetic field is therefore amplified by the field wrapping mainly.
As the field increases, $v_{\rm A}$ becomes greater.
We find that the relation (\ref{eq:mri_resol}) holds at a later time 
($\sim 0.2 \s$ for R10)
and our numerical grid can resolve, albeit marginally, 
the fastest growing mode of MRI, by which 
the field is amplified to a saturated value.
For R8, however, 
the relation (\ref{eq:mri_resol}) cannot hold so that 
the magnetic energy is possibly saturated at a lower amplitude.
We need finer numerical grids to resolve MRI for R8.

\subsection{The location of the inner boundary and the numerical grids}
\label{sec:inner-boundary}

In the present study, 
the inner boundary of the computational domain is set to be 50km,
which is larger than that in \citet{proga03} ($\sim$ 10km).
In order to examine the dependence of disk properties on 
the location of the inner boundary, $r_{\rm in}$, 
we have performed a MHD simulation of R10 with a smaller 
inner boundary of 10km and with finer 220 radial meshes
until $t \sim $ 0.8 $\s$.

\begin{figure}
 \plotone{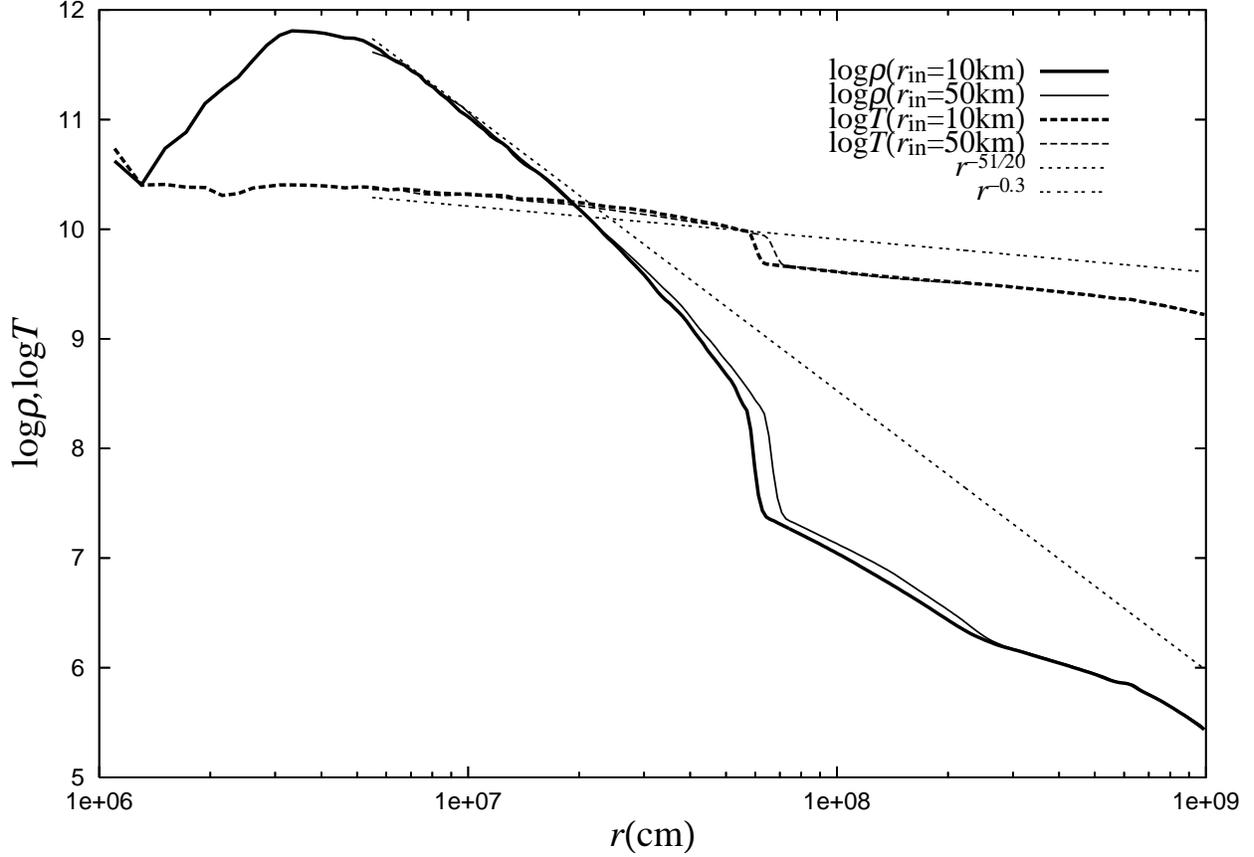}
\caption{
Radial profiles of the density (solid lines) 
and temperature (dotted lines) 
near the equatorial plane  ($\theta = 88.1^\circ$) in R10 at $0.5\s$
for $r_{\rm in} = 10$km and 50km.
Thick and thin lines correspond to the profiles 
for $r_{\rm in} = 10$km and 50km, respectively.
We also plot profiles of the density (dotted line) 
and temperature (dot dashed lines) of NDAF, 
as shown in Figure \ref{fig:r10-disk-pl}.
The profiles are independent of the location of the inner boundary.
} \label{fig:r10-disk-rin}
\end{figure}

Figure \ref{fig:r10-disk-rin} shows 
radial profiles of the density (solid lines) 
and temperature (dotted lines) 
near the equatorial plane  ($\theta = 88.1^\circ$) in R10 at $0.5 \s$
for $r_{\rm in} = 10$km and 50km.
Thick and thin lines correspond to the profiles 
for $r_{\rm in} = 10$km and 50km, respectively.
We also plot profiles of the density (dotted line) 
and temperature (dot-dashed lines) of NDAF, 
as shown in Figure \ref{fig:r10-disk-pl}.
We find that the profiles are independent of the location 
of the inner boundary.
Radial profiles of other physical quantities inside the disk 
are also independent of the location.
We also find that the disk is neutrino-cooling dominated
even if we adopt the smaller inner boundary of 10km.
The neutrino luminosity slightly increases to $1.20\times 10^{52}\ergps$ 
from $0.81\times 10^{52}\ergps$ for $r_{\rm in} = 50$km.
The luminosity has a trend of gradual decrease after the formation 
of the disk-like structure for $r_{\rm in} = 50$km (Figure \ref{fig:Lnu}).
For $r_{\rm in} = 10$km, 
the luminosity decreases to $< 1 \times 10^{52}\ergps$ after $t = 0.55 \s$
and stays roughly constant at the rate of 7-8 $\times 10^{51} \ergps$
for $t = $ 0.6-0.8 $\s$.
Furthermore, 
the disk is found to be optically thin to neutrino for $t > 0.5\s$, 
as in case with $r_{\rm in} = 50$km.

We impose the mirror symmetry on the equatorial plane, 
which is not assumed in \citet{proga03}.
If we calculate without the symmetry, 
the convective motion near the equatorial plane may be weakened 
and jets may become less collimated.
We need calculation in a computational domain 
with $0 \le \theta \le \pi$ grids to examine such the possibility.
Moreover, in light of the collapsar model, 
it may be important to investigate 
standing accretion shock instability that
is not appeared in calculation with the mirror symmetry
and whose importance has been recognized in SN explosion~\citep{blondin03}.

\subsection{Comparison between Our Models and Previous Works} \label{sec:comparison}

In MHD simulations of a 25$\Ms$ collapsar~\citep{proga03}
and a 15$\Ms$ collapsar~\citep{mizuno04a,mizuno04b}, 
a jet has been shown to be produced magnetically 
near the black hole, as in our results.
We compare our results and those obtained by 
other groups~\citep{proga03,mizuno04a,mizuno04b}.


We firstly compare disk properties of our model R10 
(Figure \ref{fig:r10-disk}) and 
those in \citet{proga03} (in their Figure 3).
We note however that  
Figure \ref{fig:r10-disk} shows the properties at $t = 1.6625 \s$
while Figure 3 in \citet{proga03} shows time-averaged properties
during $t = $ 0.2629-0.2818 $\s$.
In \citet{proga03}, 
the 25$\Ms$ collapsar is simulated with a similar distribution of 
the angular momentum and the magnetic field as in our models
after the formation 
of a black hole of 1.7$\Ms$, which is the mass of an entire iron core.
They assume non-rotating inner Si and O layers and set 
the initial distributions of velocity and density to be those of 
one-dimensional free-fall gas.
They have taken into account resistive heating and 
neutrino cooling in an optically thin regime, 
and ignored self-gravity of the star.

On the other hands, 
we have simulated the collapse of a 40 $\Ms$ star, 
whose core is assumed to collapse to a black hole promptly.
The initial distributions of physical quantities are set to be 
those of presupernova and
collapse to the black hole is mimicked with the absorption of material 
through the inner boundary.
We have ignored resistive heating but considered neutrino cooling 
with the two-stream approximation~\citep{dpn02}, 
with which we can treat the optically thin and thick regimes 
on neutrino reaction approximately.

The radial velocity for R10 is much slower than 
that in \citet{proga03} (in their top-right panel of Figure 3).
The difference is attributed to the existence of the weak shock, 
where the radial velocity sharply drops
(left bottom panel in Figure \ref{fig:r10-disk}). 
The cause of the weak shock is possibly the centrifugal barrier 
as seen in \S \ref{sec:fast-core-mhd}. 
Note that a similar accretion shock is also revealed in 
hydrodynamic calculations of the collapse of a 35$\Ms$ star~\citep{mw99}.
Disappearance of the weak shock in \citet{proga03} 
probably inherits their 
initial distributions of the inner layers above mentioned.
We note that the radial velocity inside the disk is much smaller than 
that in \citet{proga03} even if the calculation is performed 
for $r_{\rm in} = 10$ km.

The slower radial velocity leads to longer accretion time.
Thus, the cooling time via neutrino is shorter than the accretion time
in the present study (Figure \ref{fig:r10-disk-pl}), 
in contrast with \citet{proga03}.
Hence the disk is neutrino-cooling dominated in our models
while advection-dominated in their model
except for a small region inside the torus where the density reaches maximum.
We note that 
the disk structure obtained from the MHD calculation in the present study
is different from that through hydrodynamic one with
$\alpha$-viscosity~\citep{mw99}, in which 
the structure is roughly coincident with that of ADAF.

The accretion rate through the inner boundary
is estimated to be about 
$5 \times 10^{32} \g\,\s^{-1}$ ($= 0.25\DMs$) 
with fast fluctuations~\citep{proga03}, 
which is comparable with the rates 
in R8, R10, S8, and S10 after the formation of the quasi-steady disk
(see Figure \ref{fig:r10-Mdot}).
The radial profiles of the magnetic fields are similar as those of
\citet{proga03} but the magnitudes are higher.
The magnetic energy in \citet{proga03} is smaller than 
that in our models up to one order of magnitude
at the end of the calculation, 
in spite of greater initial magnetic energy;
$E_{m,\,r}$ is about $2 \times 10^{47} \ergps$
(in their Figure 1).
A jet is magnetically driven from the inner region of 
the disk in our models and \citet{proga03} 
and the jet velocity is comparable.

Moreover, we compare disk properties of our model R10 
for $r_{\rm in} = $ 10km at $t = 0.5 \s$ (Figure \ref{fig:r10-disk-rin}) 
and those in \citet{proga03} (in their Figure 3).
We find that the radial profiles of the density and temperature
in our calculation are similar to those in \citet{proga03}.
The density has a maximum of $10^{12}\rm \gpccm$ 
around $\sim$ 30-40 km.
The radial velocity rapidly increases near the inner boundary because of
strong gravity of the black hole.
The density in our run is greater than that in \citet{proga03}
due to smaller radial velocity, 
while the temperature is cooler.
In \citet{proga03}, the accretion time is shorter than 
that in our run due to greater radial velocity.
The liberated energy is therefore advected into the black hole 
with the accreting gas, which cools via neutrino inefficiently 
and becomes hotter than that in our model.
The resistive heating, which is considered in their model and 
not in our models, 
could also heat the accreting gas to higher temperature.
Consequently, 
the neutrino luminosity, which is very sensitive to the temperature
(eq. (\ref{eq:qcap})), is greater than that in our run.

Next we compare our results with those of scale-free, general
relativistic MHD simulations~\citep{mizuno04a}
for a 3$\Ms$ black hole and a density unit of $10^{10}\,\gpccm$.
They have simulated the collapse of a magnetized, rotating massive star
after the formation of the black hole 
for very short duration about $5 \, \rm ms$ 
using a simplified (gamma-law) EOS and 
very small computational domain of 360km$\times$360km.
Initial profiles of the angular velocity and the magnetic field 
are similar as in our models, 
but their model A2 corresponds to $\Omega_0 = 378\, \s^{-1}$ and 
$B_0 = 1.5 \times 10^{13}$ G, 
which is much faster rotation and greater initial field than R12.
A magnetically-driven jet is shown to emit near the hole and 
propagate along the rotational axis with a speed of up to $0.2\,c$
at $5 \, \rm ms$ much earlier than R12
because of faster rotation and greater initial magnetic field.

Our simulations are performed with the Newtonian MHD code, 
in which the pseudo-Newtonian potential is used to mimic
general relativistic effects of a black hole.
If we examine during the collapsing phase to the black hole 
or more inner region near the black hole, 
we need general relativistic MHD code as in 
\citet{mizuno04a,mizuno04b} and \citet{seki04,seki05}.
However, our numerical grid covers the region with $r \ge 50$ km, or,
5.56$r_g$ of a 3$\Ms$ black hole.
General relativistic effects are therefore small for our models.

\subsection{Implication for GRB progenitors}

Taking into account angular momentum transport 
through magnetic torque, whose importance on evolution of massive stars 
is shown in \cite{spruit02}, 
\citet{heger04,heger05} and \citet{mm04} have performed 
1.5 dimensional simulations on evolution of rotating massive stars.
They showed that
the iron-core of the stars has specific angular momentum 
up to $\sim 10^{15} \rm cm^2\,s^{-1}$ before core collapse.
It is much smaller than 
the specific angular momentum of material with the Keplerian motion 
around a 2 $\Ms$ black hole at the last stable circular orbit, 
$2 \times 10^{16} \rm cm^2\,s^{-1}$.
Consequently, single massive stars are claimed not to be a progenitor
of GRB if the magnetic braking mechanism of \citet{spruit02} operates 
in the stars~\citep{plyh05,fh05}.
Moreover, even in a binary companion star~\citep{plyh05}, 
a rapidly rotating core with the specific angular momentum of 
$2 \times 10^{16} \rm cm^2\,s^{-1}$ is unlikely to be realized 
though such core can be produced in the merger of two helium cores
during common-envelope inspiral phase of a binary system~\citep{fh05}.
It should be emphasized however that 
the angular momentum transport through magnetic torque in a massive star
is still uncertain.

Moreover, even for stars with a slowly rotating iron core, 
an accretion disk around a black hole is formed during 
the collapse of the stars 
and a jet can be produced from the disk near the hole 
if outer layers of the stars have a sufficiently high angular momentum, 
as shown in \S 3.2.
Therefore, progenitors of GRB could not always be required 
for a rapidly rotating iron core; 
The progenitors may require to 
have layers with a sufficiently high angular momentum and
a sufficiently high magnetic field.

\section{Concluding Remarks}

We have performed two-dimensional, axisymmetric MHD simulations of 
the collapse of a 40 $\Ms$, rapidly rotating star, whose 
core is assumed to collapse to a black hole promptly and 
angular momentum attains to $10^{17} \rm cm^2 \rm \, s^{-1}$, to examine 
the formation of an accretion disk around the black hole
and the jet production near the hole in the light of the collapsar model 
of GRB.
Considering two distributions of the angular velocity 
and the uniform magnetic field, whose magnitude is 
$10^8, \, 10^{10}$, and $10^{12}$ G, parallel to the rotational axis, 
inside the star before collapse, 
we investigate how angular momentum and magnetic field distributions
inside the star affect the jet production and the disk properties.
We summarize our conclusions as follows;
\begin{enumerate}
 \item After material reaches to the black hole with high angular momentum 
       of about $10^{17} \rm cm^2 \rm \, s^{-1}$, 
       a disk is formed inside a surface of weak shock, 
       which is appeared near the hole due to the centrifugal force
       and propagates outward slowly.
       The disks become a quasi-steady state for models with 
       the initial magnetic field less than $10^{10}$ G.
 \item We find that the jet can be driven by the tangled-up magnetic fields 
       even if the central core does not rotate 
       as rapidly as previously assumed and outer layers of the star has 
       a sufficiently high angular momentum.
 \item The jet is driven by the magnetic field, 
       which is dominated by the toroidal component
       and is amplified due to the wrapping of the field, 
       as long as the initial magnetic field is greater than $10^{10}$G.
       The field are chiefly amplified inside the disks and
       propagates to the polar region along the inner boundary 
       near the black hole through the Alfv{\'e}n wave.
       The jets cannot be cooled through neutrino processes.
 \item The quasi-steady disk is not an advection-dominated accretion flow 
       but a neutrino cooling-dominated disk.
       The accretion time is larger than the cooling time via neutrino.
       At an inner region ($< 100$km) of the disk, 
       the profiles of density and temperature are similar to those
       in \citet{proga03}.
 \item 
       The radial profiles of the density and the temperature
       of the quasi-steady disk are well described with those of 
       a neutrino cooling dominated disk, 
       or $\rho \sim r^{-51/20}$ and $T \sim r^{-3/10}$, 
       in which the gas pressure is dominated over the other pressure.
       These profiles are rather different from those of ADAF,
       or $\rho \sim r^{-3/2}$ and $T \sim r^{-5/8}$.
 \item Mass accretion rates in the quasi-steady disks are greater than
       $0.01 \DMs$ with large fluctuations.
       A small fraction of rest-mass energy is liberated 
       through neutrino emitted from the dense part of 
       the quasi-steady disks, which hardly trap neutrino.
       The neutrino luminosity stays at a constant level 
       of $< 8 \times 10^{51} \ergps$, 
       which is much smaller than those from a SN
       and in the previous work~\citep{proga03}.
\end{enumerate}

The mass and total energy of the jets are 0.0018-0.037 $\Ms$ and 
$5 \times 10^{49}$ - $5\times 10^{50}$ ergs, respectively (Table 2).
The jets are too heavy and weak to produce 
a relativistic fireball, or GRB, whose baryon mass and isotopic energy 
are required to be $\le 10^{-4}-10^{-5}\,\Ms$ and $10^{51}$ ergs 
\citep[e.g.,][]{hurley04}.
The jets are baryon-rich and cannot be accelerated to a
relativistic velocity, or failed GRB, 
as in the previous works~\citep{proga03,mizuno04a,mizuno04b}.
However, after the generation of the jets, 
the polar region near the black hole becomes baryon-poor and 
large magnetic field. 
If we continue to perform simulations, 
a baryon-poor outflow may be produced from the region.
We need much longer simulations and thus much wider computational domain 
to examine the ejection of these multipul jets.

Although no jet appear in R8 and S8 during the computation, 
a jet could be originated near the black hole
if the computation is performed for a longer time.
However, at the end of the computation, 
the density sharply drops near the outer boundary 
because material inside the computational domain begins to be consumed 
to infall the black hole, whose mass attains 2.64$\Ms$ for R8.
In order to perform calculation for a longer time to investigate 
whether a jet can be produced or not, we need a larger computational domain.
Our numerical grid cannot resolve
the fastest growing mode of MRI for R8 and S8.
Therefore, we need longer calculation with 
a larger computational domain and finer numerical grids
to examine the jet generation in R8 and S8.
This is our future task.

The density and temperature of jets can attain so high that 
material in the jets is in nuclear statistical equilibrium.
For relatively low density jets, 
the composition is protons and neutrons near the black hole.
As jets propagate along the rotational axis to decrease 
the density and temperature, 
alpha-rich freezeout operates inside the jets, 
which is mainly composed of \nuc{He}{4} and \nuc{Ni}{56}.
On the other hands, for dense jets, such as the jet in R12, 
the jets can be neutron-rich due to electron capture on protons,
as suggested by one-dimensional calculations 
for collapsar disk and jets~\citep{fujimoto04,fujimoto05}.
Consequently, {\itshape r}-process can operate in such dense jets 
and neutron-rich heavy nuclei, up to 3rd-peak elements can produce
inside the jets (Fujimoto et al. in preparation).

\acknowledgements{
This work was supported in part by the Japan Society for
Promotion of Science(JSPS) Research Fellowships (K.K.), 
Grants-in-Aid for the Scientific Research from the Ministry of
Education, Science and Culture of Japan (No.S14102004, No.14079202, 
No.17540267), and Grant-in-Aid for the 21st century COE program
``Holistic Research and Education Center for Physics of Self-organizing
Systems''.
We are grateful to K. Arai for his carefully reading the manuscript 
and giving useful comments.
}



\clearpage

\begin{deluxetable} {cccccccccc}
\tablewidth{0pc} 
\tablecaption{Parameters and features of models.}
\tablehead{
\colhead{model} & \colhead{$B_0$} 
& \colhead{$\Omega_0$} & \colhead{$R_0$} 
& \colhead{$t_{f}$} 
& \colhead{$|T/W|_0$} & \colhead{$|T/W|_f$} 
& \colhead{$|E_m/W|_0$} & \colhead{$|E_m/W|_f$}
& \colhead{$M_{f}$}
}
\startdata 
 R8  & $10^{8}$  & 10  & 1000 & 4.52 & 1.86 & 0.27    & 1.35e-8 & 2.99e-3 & 2.46 \nl
 R10 & $10^{10}$ & 10  & 1000 & 2.62 & 1.86 & 3.18e-3 & 1.35e-4 & 3.70e-4 & 2.63 \nl
 R12 & $10^{12}$ & 10  & 1000 & 0.36 & 1.86 & 8.48e-2 & 1.35    & 4.68e-2 & 1.88 \nl
 S8  & $10^{8}$  & 0.5 & 5000 & 3.87 & 0.11 & 0.20    & 1.35e-8 & 6.03e-3 & 3.55 \nl
 S10 & $10^{10}$ & 0.5 & 5000 & 1.34 & 0.11 & 3.98e-3 & 1.35e-4 & 9.88e-4 & 3.16 \nl
 S12 & $10^{12}$ & 0.5 & 5000 & 0.28 & 0.11 & 3.82e-2 & 1.35    & 1.01e-2 & 2.21 \nl
\enddata 
\tablecomments{
Model parameters, $B_0$, $\Omega_0$, and $R_0$, are shown in units of 
G, $\rm rad\, s^{-1}$, and km, respectively.
The calculations are stopped at the time $t_f$, 
when the mass of the central black hole is
$M_{f}$ in units of $\Ms$.
The ratios $|T/W|_{0}$, $|T/W|_{f}$, $|E_m/W|_{0}$ and $|E_m/W|_{f}$ 
are expressed in percentage.}
\end{deluxetable}

\begin{deluxetable} {cccccc}
\tablewidth{0pc} 
\tablecaption{
Jet properties
}
\tablehead{
\colhead{model} &
\colhead{$t_{\rm jet}$} &
\colhead{$M_{\rm ej}$} & 
\colhead{$(E_m)_{\rm ej}$} &
\colhead{$(E_k)_{\rm ej}$} &
\colhead{$(E_i)_{\rm ej}$}
}
\startdata
 R10 & 2.58 & 0.0010 & 2.89e-4 & 0.0274  & 0.0840 \nl
 R12 & 0.20 & 0.083  & 0.097   & 4.58    & 2.79  \nl
 S10 & 1.30 & 0.0053 & 0.045   & 0.138   & 0.606 \nl
 S12 & 0.25 & 0.033  & 0.014   & 3.57    & 5.27  \nl
\enddata
\tablecomments{
Jets are passed through 1000km at the time $t_{\rm jet}$ (s)
from the central black hole.
Ejected mass via jets, $M_{\rm ej}$, is in units of $M_\odot$.
Magnetic, kinetic, and internal energies
carried away via jets, 
$(E_m)_{\rm ej}$, $(E_k)_{\rm ej}$, and $(E_i)_{\rm ej}$, 
are expressed in $10^{50} \rm erg$.}
\end{deluxetable}
\clearpage


\end{document}